\documentclass[11pt,a4paper]{article}
\usepackage{jheppub}
\usepackage{amsmath,amsfonts,epsfig,color}
\usepackage{amssymb}
\usepackage[english, USenglish]{babel}
\usepackage{epsfig,psfrag}

\usepackage{hyperref}

\csname @addtoreset\endcsname{equation}{section}

\makeatletter
\def\timenow{\@tempcnta\time
  \@tempcntb\@tempcnta
  \divide\@tempcntb60
  \ifnum10>\@tempcntb0\fi\number\@tempcntb
  \multiply\@tempcntb60
  \advance\@tempcnta-\@tempcntb
  :\ifnum10>\@tempcnta0\fi\number\@tempcnta}
\makeatother
\def\oonoo#1#2#3{\vbox{\ialign{##\crcr
	\hfil\hfil\hfil{$#3{#1}$}\hfil\crcr\noalign{\kern1pt\nointerlineskip}
	$#3{#2}$\crcr}}}
\def\oon#1#2{\mathchoice{\oonoo{#1}{#2}{\displaystyle}}
	{\oonoo{#1}{#2}{\textstyle}}{\oonoo{#1}{#2}{\scriptstyle}}
	{\oonoo{#1}{#2}{\scriptscriptstyle}}}
\def\dt#1{\oon{\hbox{\bf .}}{#1}}  
\def\ddt#1{\oon{\hbox{\bf .\kern-1pt.}}#1}
\def\slap#1#2{\setbox0=\hbox{$#1{#2}$}
	#2\kern-\wd0{\hfuzz=1pt\hbox to\wd0{\hfil$#1{/}$\hfil}}}

\title{ Conformal anomaly of generalized form factors\\[3mm] and  finite loop  integrals}

\author[a]{Dmitry Chicherin}
\author[b,c]{and Emery Sokatchev}

\affiliation[a]{PRISMA Cluster of Excellence, Johannes Gutenberg University,
55099 Mainz, Germany}
\affiliation[b]{LAPTh, Universit\'{e} Savoie Mont Blanc, CNRS, B.P. 110,  F-74941 Annecy-le-Vieux, France}
\affiliation[c]{Theoretical Physics Department, CERN, CH -1211, Geneva 23, Switzerland}

\abstract{We reveal a new mechanism of conformal symmetry breaking at Born level. It occurs in generalized form factors with several local operators and an on-shell state of massless particles. The effect is due to hidden singularities on collinear configurations  of the momenta. This conformal anomaly is different from the  holomorphic anomaly of amplitudes. We present a number of examples in four and six dimensions. We find an application of the new conformal anomaly to finite loop momentum integrals with one or more massless  legs. The collinear region  around a massless leg creates a contact anomaly, made visible by the loop integration. The anomalous conformal Ward identity for an $\ell-$loop integral is a 2nd-order differential equation whose  right-hand side is an $(\ell-1)-$loop integral. We show several examples, { in particular the four-dimensional  scalar double box.}   }

\emailAdd{chicherin@uni-mainz.de}
\emailAdd{emeri.sokatchev@cern.ch}

\keywords{Conformal Symmetry, Anomalous Ward Identities, Form Factors, Finte Loop Integrals}

\arxivnumber{1709.03511}

\begin{document}

\renewcommand{\thefootnote}{\fnsymbol{footnote}}
%\newpage
%\pagestyle{empty}
%\setcounter{page}{0}

%%%%%%%%%%%%%%%%%%%%%%%%%%%%%%%%%%%%%

\newcommand{\norm}[1]{{\protect\normalsize{#1}}}
\newcommand{\p}[1]{(\ref{#1})}
\newcommand{\half}{{\ts \frac{1}{2}}}
\newcommand \vev [1] {\langle{#1}\rangle}
\newcommand \ket [1] {|{#1}\rangle}
\newcommand \bra [1] {\langle {#1}|}

\newcommand{\cI}{{\cal I}}
\newcommand{\cM}{{\cal M}} 
\newcommand{\cR}{{\cal R}} 
\newcommand{\cS}{{\cal S}} 
\newcommand{\cK}{{\cal K}}
\newcommand{\cL}{{\cal L}} 
\newcommand{\cF}{{\cal F}}
\newcommand{\cN}{{\cal N}}
\newcommand{\cA}{{\cal A}}
\newcommand{\cB}{{\cal B}}
\newcommand{\cG}{{\cal G}}
\newcommand{\cO}{{\cal O}}
\newcommand{\cY}{{\cal Y}}
\newcommand{\cX}{{\cal X}}
\newcommand{\cT}{{\cal T}}
\newcommand{\cW}{{\cal W}}
\newcommand{\cP}{{\cal P}}
\newcommand{\mK}{{\mathbb K}}
\newcommand{\nt}{\notag\\} 
\newcommand{\pa}{\partial}
\newcommand{\ep}{\epsilon}
\newcommand{\om}{\omega}
\newcommand{\bep}{\bar\epsilon}
\renewcommand{\a}{\alpha}
\renewcommand{\b}{\beta}
\newcommand{\g}{\gamma}
\newcommand{\s}{\sigma}
\newcommand{\la}{\lambda}
\newcommand{\tl}{\tilde\lambda}
\newcommand{\tm}{\tilde\mu}
\newcommand{\tk}{\tilde k}
\newcommand{\da}{{\dot\alpha}}
\newcommand{\db}{{\dot\beta}}
\newcommand{\dg}{{\dot\gamma}}
\newcommand{\dd}{{\dot\delta}}
\newcommand{\q}{\theta}
\newcommand{\bq}{\bar\theta}
\newcommand{\bQ}{\bar Q}
\newcommand{\tx}{\tilde{x}}
\newcommand{\tr}{\mbox{tr}}
\newcommand{\+}{{\dt+}}
\renewcommand{\-}{{\dt-}}
\newcommand{\ti}{{\textup{i}}}

\maketitle
\flushbottom

\setcounter{page}{1}\setcounter{footnote}{0}
\renewcommand{\thefootnote}{\arabic{footnote}}

\section{Introduction}

The natural observables in a CFT are the correlation functions of local gauge invariant operators. They are finite and exactly conformal functions of the coordinates, provided that the operators remain in generic  positions. Putting the operators in a singular configuration generates UV divergences and yields the breakdown of conformal symmetry. A well-known example is the lightlike limit in which the correlator becomes a Wilson loop \cite{Alday:2010zy,Eden:2010zz}. The UV divergent lightlike Wilson loops have a conformal anomaly \cite{Drummond:2007au} with interesting implications for the dual IR divergent scattering amplitudes of massless particles. Besides, even the finite tree-level amplitudes have another, collinear type of singularity, leading to a specific conformal anomaly. It was first identified and dubbed ``holomorphic anomaly" in \cite{Cachazo:2004by} and then studied in detail in \cite{Bargheer:2009qu,Korchemsky:2009hm,Bargheer:2011mm}. 

In this paper we reveal a new mechanism of conformal symmetry breaking in finite observables at the lowest, Born level of perturbation theory. They present a conformal anomaly, which is not due to divergences, but to collinear singularities in momentum space of a new type. 

These observables are generalized form factors in conformal theories in D dimensions. We discuss theories that are conformal at the classical level but not necessarily at the quantum level. The generalized form factor\footnote{The term ``generalized form factor" was introduced in  \cite{Engelund:2012re}, to distinguish it from the standard form factors involving a single operator.} involves a {time ordered} product of 
$n$ local operators, ${\cal O}(x_1) {\cal O}(x_2) \ldots {\cal O}(x_n)$. It is defined as the matrix element of this  product with an on-shell state with $m$ massless particles, $p_j^2 =0$, $j=1,\ldots,m$: 
\begin{align}\label{11}
F(x_1,\ldots,x_n|p_1,\ldots,p_m) = \vev{{\cal O}(x_1) {\cal O}(x_2) \ldots {\cal O}(x_n)| p_1,p_2,\ldots, p_m}\,.
\end{align}
The  operators naturally live in coordinate space, and the particles in momentum space, hence the mixed $x/p$ functional dependence of $F$.  One may say that the generalized form factor is a hybrid between a correlation function of local operators ($m=0$) and a scattering amplitude ($n=0$). As such, it has a much richer structure than these familiar quantities. In the present paper we are interested in the conformal properties of this new object.

We work in the Born approximation -- the lowest order of perturbation theory. At this level there are no UV or IR divergences susceptible of breaking the conformal symmetry. So it would be natural to expect that the quantities \p{11}  inherit
the classical conformal symmetry of the theory. We show that in many cases this naive believe is not true.  The action of the conformal boost transformations becomes anomalous,
\begin{align}
\left( \sum_{i=1}^n K^{(x_i)}_{\mu} + \sum_{j = 1}^m \mK^{(p_j)}_{\mu} \right) F(x_1,\ldots,x_n|p_1,\ldots,p_m) = A_\mu(x,p)\,.
\end{align}
The anomaly $A_\mu$ is a regular function, not a contact term.
Here the conformal boost generator consists of two pieces. The first piece $K^{(x)}_{\mu}$ acts in  coordinate space, e.g. for a scalar operator ${\cal O}(x)$ of  conformal dimension $\Delta$,
\begin{align} \label{boost}
K^{(x)}_{\mu;\Delta} =  i(x^2 \pa_{x^\mu} - 2 x_{\mu} x^{\nu}\pa_{x^\nu} - 2 \Delta x_{\mu})\,.
\end{align}
 The second piece $\mK^{(p)}_{\mu}$ acts in momentum space, more precisely on the lightlike momenta $p^2 =0$. For example, a 4D lightlike momentum $p_{\mu}$ factorizes in a pair of chiral and antichiral commuting helicity spinors, 
$\sigma_{\a\da}^{\mu} p_{\mu} = \lambda_{\a} \tilde\lambda_{\da}$, and the conformal boost becomes a second-order differential operator \cite{Witten:2003nn},
\begin{align}\label{1.4}
\mK^{(p)}_{\mu} = 2 \, \tilde\sigma^{\da\a}_{\mu} \frac{\pa^2}{\pa \la^{\a} \pa \tilde\la^{\da}}\,.
\end{align}
A similar realization of the conformal boost exists in six dimensions (see App.~\ref{appA}).

To be more specific, let us outline an example in 6D $\phi^3$ theory. It is conformal at the classical level, with  the scalar field having the canonical dimension  $\Delta_{\phi} = 2$.
We choose both local  operators to be elementary fields, ${\cal O}(x) = \phi(x)$, and consider
the generalized form factor with a single-particle scalar state $\phi(p)$ in the Born approximation,
\begin{align} \label{25}
F(x_1,x_2|p) = \vev{{\cal O}(x_1) {\cal O}(x_2)| \phi(p)}_{\rm Born}= \frac{g}{x_{12}^2} \frac{e^{i p x_1 } - e^{i p x_2 }}{i(p x_{12})}\,.
\end{align}
This expression is manifestly translation, Lorentz and dilatation invariant, but what about conformal boosts?
This symmetry is broken, as shown  by the anomalous Ward identity 
\begin{align} \label{WIintr}
\left( \sum_{i=1}^2 K^{(x_i)}_{\mu;\Delta=2} + {\mK}^{(p)}_{\mu}\right) \,  F(x_1,x_2|p) = 
 -g\, p_\mu \,  \int^1_0 d\xi \,  \xi\bar\xi\, e^{ i (p x_{1})\xi +i (p x_2) \bar \xi} \,, \quad \bar\xi :=1-\xi \,.
\end{align}
We can Fourier transform the local operators of the generalized form factor \p{25} from position to momentum space, $x_1,x_2 \to q_1, q_2$,
\begin{align}\label{19}
\tilde F(q_1,q_2|p) = \vev{{\cal O}(q_1) {\cal O}(q_2)| \phi(p)}_{\rm Born} = \frac{g}{q_1^2 q_2^2}\delta^{(6)}(q_1 + q_2 +p) \,.
\end{align}
In  momentum space the anomaly in the Ward identity \p{WIintr} becomes a {\it contact term},
\begin{align}\label{mWI}
K^{(q,p)}_\mu \, \tilde F(q_1,q_2|p) =4\pi^3 g \, p_{\mu}\int_0^1 d\xi\, \xi\bar\xi\, \delta^{(6)}(q_1 + \xi p) \, \delta^{(6)}(q_2 +\bar\xi p)\,.
\end{align}
The conformal anomaly arises on a configuration where the off-shell momenta of the local operators become collinear with the on-shell momentum of the  particle, $q_1^\mu \sim q_2^\mu \sim p^\mu$. 

The collinear `holomorphic'  anomaly  of scattering amplitudes mentioned above has a different origin. Consider, e.g., the MHV  tree-level color ordered 4D amplitude for $n$ gluons of helictites  $(--+\ldots+)$:
\begin{align}\label{111}
\cA_n \sim \frac{\vev{12}^3\, \delta^{(4)}\left(\sum_{i=1}^n \la_i\, \tl_i \right)}{\vev{23} \vev{34} \ldots \vev{n1}}\,,
\end{align}
where $\vev{ij} := \la^\a_i \la_{j\, \a}$ are Lorentz invariant contractions of the chiral helicity spinors. This amplitude   has complex poles at $\la_i \sim \la_{i+1}$, i.e. where the momenta of two adjacent particles (except the first two) become collinear, $p_i \sim p_{i+1}$. When the derivative $\pa/\pa\tl_i$ from the conformal boost generator \p{1.4} hits such a pole,  it produces a contact term $\sim \delta^{(2)}(\vev{i\, i+1})$. This resulting anomaly relates  $\mK_{\mu} \cA_n$ to $\cA_{n-1}$.

The mechanism of our new anomaly is more subtle, due to the presence of off-shell momenta in the problem. Looking at the expression \p{19}, it is hard to detect an obvious problem, like the complex poles in \p{111}, that is susceptible of breaking the symmetry. In reality, the origin of the anomaly \p{mWI} is a hidden singularity in the product of the two scalar propagators in \p{19} (for details see Sect.~\ref{s2}). Importantly, the anomaly only takes place in generalized form factors, i.e. when more than one off-shell momenta are involved. Indeed, the same 6D $\phi^3$ vertex can give rise to a standard form factor with one operator and two massless particles, 
\begin{align}\label{}
\tilde F(q|p_1,p_2) = \vev{\cO(q)|\phi(p_1) \phi(p_2)} _{\rm Born}= \frac{g}{q^2}\, \delta^{(6)}(q+p_1+p_2)\,.
\end{align}
Its conformal symmetry is not broken.  

In the present paper we discuss several other examples of the same phenomenon. In Sect.~\ref{s31} we examine the analog of  the generalized form factor \p{25}, based on the 4D vertex $\phi^4$. In Sect.~\ref{s32} we use a 4D vertex of the Yukawa type as an example involving fermion operators with spin.  In Sect.~\ref{s33} we consider a gauge theory coupled to scalar matter. Here the generalized form factor exhibits both types of anomalies, the familiar holomorphic one and the new collinear anomaly involving the off-shell momenta. In all of these cases the anomaly is due to a hidden singularity on a collinear momentum configuration. However,  revealing this singularity directly in momentum space is hard. Instead, we study the anomalies in the mixed coordinate-momentum representation like \p{25}. The Fourier transform to position space smears the contact anomaly and makes it easily detectable. Its explicit form is obtained most efficiently  by the method of Lagrangian insertion, inspired by the treatment of the conformal anomaly of the lightlike Wilson loop \cite{Drummond:2007au}.

We  emphasize that our examples are not supersymmetric, although some of them can easily be extended to superconformal theories. We are concerned with the breakdown of ordinary conformal symmetry.  In this context we should mention the papers \cite{Korchemsky:2009hm} and \cite{CaronHuot:2011kk}, where the anomaly of the dual $\bar Q$ supersymmetry of the $\cN=4$ SYM superamplitudes is interpreted as originating from collinear singularities. This dual supersymmetry is equivalent to ordinary $S$ superconformal symmetry, and hence the $\bar Q$ anomaly implies a conformal anomaly as well. It should however be pointed out that $\bar Q$ supersymmetry is an on-shell symmetry, realized non-linearly on the chiral superamplitudes or on the dual super-Wilson loops, see \cite{Bullimore:2011kg,Chicherin:2016fac,Chicherin:2016fbj}. Our conformal anomaly is much more basic, it has to do with a standard linear symmetry. 

One might think that the conformal anomaly \p{mWI} is almost invisible due to its contact nature in  momentum space. In reality, it has an interesting non-trivial manifestation for loop integrals. The general belief is that conformal symmetry breaking at the quantum level
is related to divergences of loop integrals and that finite quantum corrections could not spoil the symmetry. Here we show that this is not true. The contact conformal anomaly of the trivalent vertex \p{mWI} serves as a `seed' that, being inserted in a naively conformal loop integral, localizes one of the loop integration and produces a regular contribution which breaks the conformal symmetry. The corresponding anomaly is not contact and is easily detectable.   The anomaly occurs in various finite 6D and 4D integrals with one or more legs on the massless shell. The insertion of the 6D vertex \p{mWI} or of its 4D analog reduces the transcedentality weight. Thus, the anomalies of the 6D one-loop boxes (Sect.~\ref{s3}) and hexagon (Sect.~\ref{s42}) are simply given by logs and rational factors; that of the 4D double box  (Sect.~\ref{secdblbox}) by dilogs, etc. In this way we can find 2nd-order differential equations for such integrals, with an easily predictable right-hand side. This procedure might provide us with useful information about the double box and other unknown integrals.

The paper ends with several technical appendices. In App.~\ref{appA} we summarize the realization of the conformal group in position and momentum spaces, including lightlike momenta. In App.~\ref{appB} we present a direct proof of the anomaly \p{WIintr}. In App.~\ref{s321} we give the derivation of the anomaly in a gauge theory. In App.~\ref{appD} we discuss the cuts (discontinuities) of the 6D box integrals and of the corresponding conformal Ward identities.

\section{Generalized form factor in $D=6$ scalar $\phi^3$ theory}\label{s2}

In this section we disscuss in detail the simple example from the Introduction, that of a would-be conformal form factor in 6D scalar $\phi^3$ theory with Lagrangian $L = \frac12 (\pa_\mu \phi )^2 + \frac{g}{3!} \phi^3$. We show that the careful treatment of the singularities exhibits a conformal  anomaly. 

\subsection{The $\phi^3$ vertex as a generalized form factor}

Let us start with the Born-level three-point Green's function
\begin{align}\label{2.1}
 \vev{\phi(x_1) \phi(x_2) \phi(x_3)}_g = \frac{g}{i \pi^3}\, \int  \frac{d^6 x_0}{x_{10}^4 x_{20}^4 x_{30}^4}\,.
\end{align}
Here we use the free 6D massless scalar propagator $1/x^4$.\footnote{In this paper the massless scalar propagator in $D-$dimensional  momentum space is defined with Minkowski  signature $(+-\ldots-)$ and Feynman prescription $1/(q^2+i\ep)$. In coordinate space  it becomes $e^{-i\pi (D-1)/2} 2^{D-2}  \Gamma(D/2-1) \pi^{D/2} (x^2-i\ep)^{1-D/2}$.}
The integral  \p{2.1} is finite and manifestly conformally covariant.\footnote{Conformal symmetry fixes the form of the three-point function up to a normalization constant, $C(x^2_{12} x^2_{13} x^2_{23})^{-1}$. We do not need this explicit expression for our argument.} This is natural, since the classical theory is conformal. We stay at Born level, so the non-vanishing $\b-$function plays no role. 

Now, let us define the generalized form factor obtained by amputating one leg of the three-point function. To this end we first Fourier transform, e.g., point $x_3$, 
\begin{align} \label{2.2}
\int \frac{d^6 x_0}{i \pi^3} \frac{1}{x_{10}^4 x_{20}^4 x_{30}^4} \ \rightarrow\  \int \frac{d^6 x_0}{i \pi^3} \frac{e^{i p x_0 }}{x_{10}^4 x_{20}^4\, p^2 } =: I(x_1,x_2,p) \,,
\end{align} 
then multiply by $p^2$ and put the leg on shell,
\begin{align}\label{Fx}
F(x_1,x_2,p):=\vev{\phi(x_1) \phi(x_2)|\phi(p)}_{g}  = 
g\, \lim_{p^2\to0} \, \int \frac{d^6 x_0}{i \pi^3} \frac{e^{i p x_0 }}{x_{10}^4 x_{20}^4}\,.
\end{align}
We would expect the result to be conformal as well because the amputation procedure does not involve any IR or UV divergences. In fact, this is not true, as shown below and has  already been announced in the Introduction.
The breakdown of conformal symmetry  is described by the anomalous Ward identity \p{WIintr} in the mixed $x/p$ space or equivalently, by \p{mWI} in momentum space. Where does the anomaly come from?

The Fourier transform of the three-point function \p{2.2} is conformal for $p^2 \neq 0$,
\begin{align} \label{Ixp'}
\left( \sum_{i=1}^2 K^{(x_i)}_{\mu;\Delta=2}  + K^{(p)}_{\mu;\Delta=2} \right) I(x_1,x_2,p) = 0 \,.
\end{align}
We want to understand what happens when $p^2 = 0$, that is, with the form factor \p{Fx}. Its explicit expression shown in \p{25}  is worked out in App.~\ref{appB2} (see \p{b8}).
The conformal properties of the result are also examined in App.~\ref{appB2}. We apply the off-shell $x-$space generator $K^{(x_i)}_{\mu;\Delta=2}$ (see \p{a2}) and the on-shell $p-$space generator $\mK^{(p)}_{\mu}$ (see \p{a12}). The calculation yields the {anomalous Ward identity} (see \p{b9})  
\begin{align}\label{26}
\left( \sum_{i=1}^2 K^{(x_i)}_{\mu;\Delta=2} + {\mK}^{(p)}_{\mu}\right) \, F  = p_\mu\,   A(x_1,x_2,p) \, ,
\end{align}
where the  anomaly function $A(x_1,x_2,p)$  can be written in the integral form (see \p{b17'})\footnote{From here on we omit the coupling constant $g$.}
\begin{align} \label{27}
A(x_1,x_2,p) =  - \int^1_0 d\xi \,  \xi\bar\xi\, e^{ i (p x_{1})\xi +i (p x_2) \bar \xi}    \,.
\end{align}
Moreover, as shown  in App.~\ref{appB2}, there exists no function made of the two points $x_1,x_2$ and the {\it lightlike} momentum $p$ which satisfies the {\it exact} conformal Ward identity \p{Ixp'}. This is only possible if $p^2\neq0$.

What is the deep reason for this anomaly? The Fourier integral in \p{2.2} comes from the conformal three-point function \p{2.1}, so it is invariant under the combined action of the conformal boosts $K^{(x_i)}_{\mu;\Delta=2}$ and $K^{(p)}_{\mu;\Delta=2}$. In particular, the factor $e^{ipx_0}/p^2$ transforms with a weight factor $\sim x_{0\, \mu}$ needed to compensate the weight of the measure and of the two $x-$space propagators. After the amputation in \p{Fx} we act with $K^{(p)}_{\mu;\Delta=4}$, to adjust for the weight of the missing propagator factor. One would think that the integral should remain invariant. However, the weight factor $\sim x_{0\, \mu}$ together with the on-shell condition $p^2=0$ make the integral diverge. To regularize it, we may modify the dimension of the measure, $d^{6-2\ep}x_0$, but this creates a mismatch of the conformal weights $\sim\ep$. This factor multiples the pole $1/\ep$ of the divergent integral and results in a finite anomaly term. 

We exploit this mechanism in Sect.~\ref{s23} for the alternative, and  in practice most efficient proof of the anomalous Ward identity. It is inspired by the treatment of the conformal anomaly of the lightlike Wilson loop in Ref.~\cite{Drummond:2007au} and consists in inserting the Lagrangian, $x^\mu_0 L(x_0)$,  in the path integral as a way of revealing the anomaly. For the third proof in App.~\ref{appB3}, we start with the off-shell Fourier integral \p{2.2}, then act with the conformal generator and take the on-shell limit.

\subsection{Anomalous Ward identity in momentum space}\label{s2.3}

To elucidate the nature of the anomaly, we Fourier transform the form factor \p{Fx} from position to momentum space,  $x_1, x_2 \to q_1, q_2$, and obtain (up to a normalization factor)
\begin{align}\label{114}
\tilde F(q_1, q_2, p) = \frac{\delta^{(6)}(q_1 + q_2 + p)}{q_1^2 q^2_2}\,.
\end{align}
We want to study its behavior under the off- and on-shell conformal boosts \p{Kq} and \p{a12}, respectively. 
At first sight, apart form the singularities at $q^2=0$ (regulated with the $i\ep$ prescription), this distribution shows no particular problem which might cause the anomaly \p{mWI}.   However, there is a hidden singularity in the collinear regime $p \sim q_1 \sim q_2$. 

This is difficult to see directly in momentum space, therefore we start form the mixed $x/p$ representation of the form factor \p{25} and of its anomaly \p{26}. Fourier transforming both sides of eq.~\p{26} with the anomaly in the form \p{27}, we  obtain the {anomalous conformal Ward identity} for the form factor in momentum space
\begin{align}\label{115}
&\left(  \sum_{i = 1}^2 K^{(q_i)}_{\mu;\Delta=2} + {\mK}^{(p)}_{\mu} \right)  \,
\tilde F   = 4 i \pi^3 \, p_{\mu}\, \delta^{(6)}(q_1 + q_2 +  p)\int^1_0 d \xi\, \xi\bar\xi\, \delta^{(6)}(q_1 + \xi p)\,.
\end{align}
The anomaly is localized on a configuration where the three momenta become collinear,
\begin{align}\label{2.12}
q_1 = -\xi p\,, \qquad q_2 = - (1-\xi) p\,, \qquad 0\leq\xi\leq 1  \,.
\end{align}

According to App.~\ref{appA2}, the conformal boost goes through the momentum conservation delta function, so the Ward identity  \p{115} can be rewritten in the simplified form
\begin{align}\label{218}
&\left( K^{(q)}_{\mu;\Delta=2} + {\mK}^{(p)}_{\mu} \right) 
\frac{1}{q^2\, (q+p)^2 }    = 4 i \pi^3 \, p_{\mu}\int^1_0 d \xi\, \xi\bar\xi\, \delta^{(6)}(q + \xi p) =: A^{(6D)}_{\mu}(p;q) \,.
\end{align}
We use this result in Sects.~\ref{s3} and \ref{s42} to derive the conformal Ward identities for the 6D box and hexagon integrals.

We emphasize that the anomaly is due to the  on-shell leg (or massless particle) in the form factor. Indeed, in momentum space the three-point function \p{2.1} has the form
 \begin{align}\label{}
\vev{\phi(q_1) \phi(q_2) \phi(q_2)}  = \frac{\delta^{(6)}(q_1+q_2+q_3)}{q_1^2\, q_2^2\, q_3^2}\,.
\end{align}
Being the Fourier transform of the exactly conformal integral \p{2.1}, this distribution satisfies an anomaly-free conformal Ward identity. When the third leg is amputated,  the distribution develops a collinear singularity on the surface \p{2.12}, which yields the anomaly. 

We can apply the same argument to the 6D form factor $\vev{\cO(x) | \phi(p_1) \phi(p_2)}$ with a single operator and two on-shell legs. Its expression  is  
\begin{align}\label{213}
\vev{\cO(x) | \phi(p_1) \phi(p_2)}   = \frac{e^{i (p_1+p_2) x}}{(p_1+p_2)^2 +i \ep} \,,
\end{align}
or in momentum space
\begin{align}\label{222}
\vev{\cO(q) | \phi(p_1) \phi(p_2)}  = \frac{\delta^{(6)}(q + p_1+p_2)}{q^2+i\ep }\,.
\end{align}
The collinear singularity at $p_1 \sim p_2$ is regularized by the $i\ep$ prescription. Acting with the conformal boost generators on the right-hand side of \p{213}, we find zero.  Unlike the holomorphic anomaly of the scattering amplitudes discussed in the Introduction, here the collinear regime $p_1 \sim p_2$ does not yield a breakdown of conformal symmetry. 

This example illustrates the general phenomenon of collinear conformal anomaly. In addition, it also serves as a `seed' for revealing the conformal anomaly of some finite loop integrals discussed in Sections~\ref{s3} and \ref{s42}.

\subsection{Derivation of the conformal anomaly by Lagrangian insertion}\label{s23}

The method used in App.~\ref{appB} is not easy to generalize to form factors with several operators $\cO(x)$. The corresponding expression (the analog of  eq.~\p{25}) depends on many kinematical variables. Working it out  and finding its anomaly is a non-trivial task.

The most efficient way of deriving the conformal anomaly is by a  Lagrangian insertion in the  path integral \cite{Drummond:2007au}.
We consider the Green's function $\vev{\phi(x_1) \phi(x_2) |\phi(p)}$ in the theory with modified action
\begin{align}\label{}
    S= \int \,    \frac{d^D x}{g^2 \mu^{2\ep}}\, \left( \frac12\pa^\mu\phi \pa_\mu\phi + \frac1{3!}\phi^3 \right) . 
    \end{align}
Here the dimension of the measure has been changed to $D=6-2\ep$ while the scalar field keeps its canonical dimension $\Delta_{\phi} = 2$ and the difference is compensated by the dimensional regularization scale $\mu$. When performing a conformal transformation with generator $K^\mu$ in the path integral, there is a mismatch between the canonical dimension $\Delta_L=6$ of the Lagrangian $L(x_0)$ and the modified measure $\int d^Dx_0$. This leads to a breakdown of conformal invariance  originating from the  term $\Delta\, x^\mu_0 =(D-\Delta_L) \, x_{0}^\mu =-2 \ep x_{0}^\mu $ in the conformal boost \p{a2}. The symmetry breaking term takes the form of an insertion $\sim \ep   \int d^D x_0\,  x^\mu_0\, L(x_0)$ into the Green's function. Then, {if the integral  over $x_0$ has a pole $1/\ep$,  in the limit $\ep\to0$ this results in a finite conformal anomaly.} Naively, we would not expect a pole in a tree-level calculation, but once again this is not true.

So, to obtain the anomaly at lightlike $p^2 = 0$ we need to calculate 
\begin{align} \label{Lins6D}
\lim_{\ep \to0}\,  (-2i \ep) \int d^D x_0\,  \vev{\phi(x_1) \phi(x_2) \left(\frac{x_0^\mu}{g^2 \mu^{2\ep}}L(x_0)\right)|\phi(p)}\,.
\end{align}
We first consider the  insertion of the cubic term in the Lagrangian at $p^2 \neq 0$,\footnote{Alternatively, we  could insert the kinetic term $L_{\mathrm{kin}} =\phi \Box \phi$ into the scalar propagator lines. It is easy to see that this is equivalent to inserting the interaction term.} 
\begin{align} \label{intphiphiL}
&\int d^D x_0\vev{\phi(x_1) \phi(x_2) L(x_0)|\phi(p)}_{p^2\neq0} = 
\int \frac{d^D x_0}{i \pi^\frac{D}{2}} \frac{e^{i p x_0 }}{x_{10}^4 x_{20}^4} =: I(x_1,x_2,p)\,.
\end{align}
Here we omitted the coupling constant and the dimreg scale $g^2 \mu^{2\ep}$. We are allowed to use the $(D=6)-$dimensional form $1/x^4$ of the $x$-space propagators, since the finite $O(\ep^0)$ part of this integral is irrelevant for our purposes.

We go through the standard procedure of introducing Schwinger parameters in the integral in eq.~\p{intphiphiL} and  
doing the space-time integrations, with the result
\begin{align}
I(x_1,x_2,p) 
&= - \int^1_0 d\xi \int^\infty_0 d\eta\, \eta^{3-\tfrac{D}{2}}\, \xi\bar\xi \, 
e^{-\frac{p^2}{4\eta} - \xi \bar\xi \eta x_{12}^2 + i (p x_{1})\xi + i (p x_{2})\bar\xi}\ .  \label{2.4}
\end{align}
Then, to produce the factor $x^\mu_0$ in \p{Lins6D}, we differentiate with respect to $p_\mu$  and set $p^2 =0$:\footnote{The result \p{2.4} is obtained by Wick rotation to Euclidean space. Before taking the limit $p^2\to0$ we rotate back to Minkowski space, keeping $p^2>0$ and $x^2_{12}>0$ for convergence. The final result \p{2.20} can be analytically continued to  all values of $x^2_{12}$.}
\begin{align}\label{2.19}
&-i \int d^D x_0 \,x_0^{\mu}\vev{\phi(x_1) \phi(x_2) L(x_0)|\phi(p)}_{p^2=0} \notag\\
&= \int^1_0 d\xi \int^\infty_0 d\eta\, \eta^{3-\tfrac{D}{2}}\, \xi\bar\xi\, \left( -\frac{p^\mu}{2\eta} + i \xi x_{12}^\mu\right)
e^{- \xi \bar\xi \eta x_{12}^2 + i (p x_{12})\xi + i (p x_2)}\,.
\end{align}
We are interested in the pole part of this expression, so only the first term in the parentheses is relevant. We find
\begin{align}\label{2.20}
&- i \int d^D x_0 \,x_0^{\mu}\vev{\phi(x_1) \phi(x_2) L(x_0)|\phi(p)} =
\frac{p^\mu}{2\ep} A(x_1,x_2,p) +O(\ep^0) \,,
\end{align}
with $A(x_1,x_2,p)$ defined in \p{27}. Substituting this result in \p{Lins6D} reproduces the {anomalous conformal Ward identity} \p{26}. 

It should be noted that the same mechanism of Lagrangian insertion can measure the dilatation anomaly, if existing. In our example $\int d^D x_0\, L(x_0) \ldots$ produces the integral \p{2.4} which does not have a pole at $p^2 = 0$ and hence {there is no anomalous dimension}. So, the anomaly is only in the conformal boost. The reason for the pole there is the insertion of $x^\mu_0$ which makes the integral in \p{Lins6D} diverge. 

In conclusion, this method of deriving the conformal anomaly is the most efficient one. The contact term in \p{115} is hard to detect directly. The Fourier transform $ (q_1,q_2) \to (x_1,x_2)$ smears the contact anomaly and makes it easily visible in the form \p{26}, \p{27}.

\section{Conformal anomalies in $D=4$ theories}\label{se3}

In this Section we show three examples of anomalous conformal Ward identities for tree-level generalized form factors in 4D conformal theories. 
Firstly, we study the analog of the 6D form factor \p{Fx} with a single-particle state for the 4D $\phi^4$ theory. Secondly, we examine a Yukawa vertex where the two fermions are treated as operators and the scalar is the on-shell particle. Finally, 
we consider a Yang-Mills field coupled to a scalar and give an example of a generalized form factor of two composite operators with a three-particle state. The Ward identities are derived in the mixed $x/p$ representation by the method of Lagrangian insertion.  The anomaly in the momentum $q/p$ representation occurs when the off-shell momenta $q$ become collinear with the on-shell ones $p$. 

\subsection{Scalar $\phi^4$ theory} \label{s31}

\subsubsection{The $\phi^4$ vertex as a generalized form factor}

The 4D analog of the 6D `seed'  form factor from Section~\ref{s2} is defined by the $\phi^4$ vertex
\begin{align}\label{8.45} 
F(x_1,x_2,x_3,p) := \vev{\phi(x_1) \phi(x_2) \phi(x_3)|\phi(p)}_{\rm Born} 
=\left. \int \frac{d^4 x_0}{i \pi^2} \frac{e^{i p x_0 } }{x_{10}^2 x_{20}^2 x_{30}^2}\ \right\vert_{p^2 = 0}\,.
\end{align}
The $\phi^4$ theory is conformal at tree level, so naively the generalized form factor \p{8.45} should be conformally invariant. However, this is not true as we show below.

Unlike the 6D analog in \p{25}, this form factor is not expressible in elementary functions in the mixed $x/p$ representation.
Introducing Schwinger parameters and doing the space-time integrations we find the following integral expression  
\begin{align} \label{FPint}
&F(x_1,x_2,x_3,p)   = \int d \Omega(\a,\b,\gamma) \,
e^{i \alpha (p x_1) +i \beta (p x_2) + i \gamma (p x_3)} \, \Lambda^{-1}(\a,\b,\gamma,x_1,x_2,x_3) \,,
\end{align}
where the measure is $d \Omega(\a,\b,\gamma):= d \alpha \, d \beta \, d \gamma \, \delta(\alpha + \beta +\gamma - 1)$ and we integrate over $\alpha, \beta, \gamma \geq 0$. We also use the shorthand notations $\bar \alpha :=  1 - \alpha$, etc., as well as
\begin{align} \label{La}
\Lambda:= \alpha \bar\alpha x_1^2 + \beta \bar\beta x_2^2 + \gamma \bar\gamma x_3^2 
- 2 \alpha\beta (x_1 x_2) - 2 \alpha\gamma (x_1 x_3) - 2 \beta\gamma (x_2 x_3)\,.
\end{align}

\subsubsection{Conformal anomaly in $x/p$ space}

Here we show that the generalized form factor \p{8.45} satisfies the conformal Ward identity  
\begin{align}\label{8.48}
\left(  \sum_{i = 1}^3 K^{(x_i)}_{\mu;\Delta=1} + {\mK}^{(p)}_{\mu} \right)  F(x_1,x_2,x_3,p)  =
p_{\mu} \, A(x_1,x_2,x_3,p)
\end{align}
with the anomaly given by  
\begin{align} \label{hphi4}
A(x_1,x_2,x_3,p) :=  -\int d \Omega(\a,\b,\gamma) \,
e^{i \alpha (p x_1) +i \beta (p x_2) + i \gamma (p x_3)}\,,
\end{align}
or explicitly, 
\begin{align} \label{36}
& A(x_1,x_2,x_3,p) =  \sum_{\sigma \in \mathbb{Z}_3} \frac{e^{i\, p\, x_{\sigma_1}}}{(x_{\sigma_1 \sigma_2}p)(x_{\sigma_1 \sigma_3}p)}\,.
\end{align}

We apply the method of  Lagrangian insertion in $D=4-2\ep$ dimensions, described in Section~\ref{s23}. To obtain the conformal boost variation of the generalized form factor $F$ \p{8.45} we need to calculate
the residue
\begin{align} \label{Lins4D}
\lim_{\ep \to0}\,  (-2i \ep) \int d^D x_0\, x_0^\mu \, \vev{\phi(x_1) \phi(x_2)  \phi(x_3) L(x_0) |\phi(p)}\,,
\end{align}
where $p^2 = 0$ and $L(x_0) \sim \phi^4(x_0)$. We start with the integrated Lagrangian insertion in the off-shell version of $F$ given by
\begin{align}
I(x_1,x_2,x_3,p) := \int \frac{d^D x_0}{i \pi^\frac{D}{2}}\, \frac{e^{i p x_0 }}{x_{10}^2 x_{20}^2 x_{30}^2}\quad {\rm with} \ p^2\neq 0\,.
\end{align}
It has a Schwinger parameter representation similar to \p{FPint}, 
\begin{align}\label{4.2}
&I(x_1,x_2,x_3,p) = \int d\Omega(\alpha,\beta,\gamma) \,
e^{i \alpha (p x_1) +i \beta (p x_2) + i \gamma (p x_3)} \int_0^{\infty} d \eta \, \eta^{\epsilon}\, \exp\left[ -\frac{p^2}{4\eta} - \eta \Lambda \right].
\end{align}
Then we differentiate  the  integral \p{4.2}
with respect to $p_{\mu}$, set $p^2 = 0$ and extract the pole:   
\begin{align} 
&\pa_{p_{\mu}} I|_{p^2 = 0} = - \frac{p^\mu}{2} \int d \Omega(\alpha , \beta , \gamma)\, e^{i \alpha (p x_1) +i \beta (p x_2) + i \gamma (p x_3)} \int_0^{\infty} d \eta \, \eta^{\epsilon-1}\, e^{-\eta \Lambda} + O(\ep^0) \notag \\
& =   \frac{p^\mu}{2\ep} \, A(x_1,x_2,x_3,p) + O(\ep^0)\,,
\end{align}
with $A$ defined in \p{hphi4}. 
Inserting the pole in \p{Lins4D} we arrive at  the Ward identity \p{8.48}. 

We have confirmed this result by a direct check, namely, we acted with the conformal boost generators on the integrand of \p{FPint}, then numerically integrated over $\alpha,\beta,\gamma$ and compared with the right-hand side of \p{8.48}.

\subsubsection{Conformal anomaly in $q/p$ space}

The Fourier transform $x_i\to q_i$ in \p{8.45} defines the generalized form factor  
in momentum space,
\begin{align}\label{851}
\tilde F(q_1,q_2,q_3,p) = \frac{1}{q^2_1 q^2_2 q^2_3}\ \delta^{(4)}(P)  \,,
\end{align}
where $P = q_1 +q_2 + q_3+p$ is the total momentum and the $i\ep$ prescription is implied. Its 
conformal anomaly is the Fourier transform of \p{8.48}, \p{hphi4}:  
\begin{align}\label{850} 
\left(  \sum_{i = 1}^3 K^{(q_i)}_{\mu;\Delta=1} + {\mK}^{(p)}_{\mu} \right)   \tilde F &= 4  \pi^4 p_\mu \delta^{(4)}(P) \int d \Omega(\a,\b,\gamma)  \, \delta^{(4)}(q_1 + \alpha p) 
\, \delta^{(4)}(q_2 + \beta p)\,. 
\end{align}
Like in the 6D case \p{115}, the anomaly has support on the kinematic configuration where the three (off-shell) momenta associated with the `operators' $\phi(x_i)$ become collinear with the on-shell momentum $p$ of the incoming massless particle,
\begin{align}
q_1 = - \alpha p \,,\quad  q_2 = - \beta p \,,\quad q_3 = (\alpha + \beta - 1) p\,.
\end{align}

According to App.~\ref{appA2}, we can omit  one of the off-shell momenta and the momentum conservation delta function in \p{850},  
\begin{align}\label{314} 
&\left(  \sum_{i = 1}^2 K^{(q_i)}_{\mu;\Delta=1} + {\mK}^{(p)}_{\mu} \right) \frac{1}{q_1^2 q_2^2 (q_1+q_2+p)^2} \nt 
&= 4\pi^4 p_\mu \int d \Omega(\a,\b,\gamma)  \, \delta^{(4)}(q_1 + \alpha p) \, \delta^{(4)}(q_2 + \beta p) =: A_{\mu}^{(4D)}(p;q_1,q_2)\,.
\end{align}
We use this result in Section \ref{secdblbox} to derive the conformal Ward identity for the 4D six-leg double box integral.

\subsection{Yukawa vertex}\label{s32}

In this subsection we show an example of a generalized form factor for operators with Lorentz spin. The 4D Yukawa-type vertex $\int d^4 x\,  \psi_\a(x) \psi^\b(x)\phi(x)$ is conformally covariant.  We consider the following form factor
corresponding to this trivalent vertex,
\begin{align}\label{FYuk}
F{}_{\a}{}^{\b}(x_1,x_2,p) := \vev{\psi_{\a}(x_1) \psi^{\beta}(x_2) |\phi(p)}_{\rm Born}
=  (\pa_{x_1} \tilde\pa_{x_2})_{\a}{}^{\b} \, I(x_1,x_2,p)_{p^2 = 0,\ep \to 0}\,.
\end{align}
It is given by the double derivative of the scalar integral
\begin{align}\label{ID}
I(x_1,x_2,p) := \int \frac{d^{D} x_0}{i \pi^{\frac{D}{2}}} \frac{e^{i p x_0}}{ x_{10}^2\, x_{20}^2 }\,,
\end{align}
where $D=4-2\ep$ and  $\ep>0$ is an intermediate regulator.
Evaluating the integral yields a compact expression for the form factor \p{FYuk},
\begin{align}\label{Yukawa4}
F{}_{\a}{}^{\b}(x_1,x_2,p) &= - \lim_{\ep\to0}\, (\pa_{x_1} \tilde\pa_{x_2})_{\a}{}^{\b}\  \Gamma(\ep )\, (x_{12}^2)^{-\ep}\,  \int^1_0 d \xi \,
e^{i \xi (p x_1) + i \bar \xi (p x_2)}\nt
&= \frac{2}{x_{12}^2} \int^1_0 d \xi \left[ i (x_{12} \tilde p)_{\a}{}^{\b} - 2 \left(1+ i (p \tilde x_{12}) \xi \right)\delta_{\a}{}^{\b}  \right]
e^{i \xi (p x_1) + i \bar \xi (p x_2)} \,.
\end{align}

Like in the preceding sections, we expect the anomalous conformal Ward identity 
\begin{align} \label{WardYuk}
\left(  \sum_{i = 1}^2 K^{(x_i)}_{\mu;\Delta=3/2} + {\mK}^{(p)}_{\mu} \right)  F{}_{\a}{}^{\b}(x_1, x_2,p) = 
p_{\mu}\, A{}_{\a}{}^{\b}(x_1,x_2,p)\,.
\end{align}
This is a matrix relation because the form factor $F$ carries (chiral) Lorentz indices. The space-time conformal boost generator \p{a2} involves a term acting on them, $ - 2 i x^{\nu} \Sigma_{\nu \mu}$. In the case at hand $\Sigma_{\mu\nu} = \frac{i}{4}(\sigma_{\mu}\tilde\sigma_{\nu}-\sigma_{\nu}\tilde\sigma_{\mu})$ is the Lorentz generator in the spinor representation $(1/2,0)$.
The conformal weights at points $1$ and $2$ equal $3/2$ as for a Dirac spinor field.

The conformal anomaly $A$ is obtained by the method of Lagrangian insertion  from Section \ref{s23}. We need to calculate the  residue 
\begin{align} \label{hlim}
p_{\mu\,} A_{\a}{}^{\b}(x_1,x_2,p)  =  \lim_{\ep \to0} 2 \ep  \, \pa_{p^\mu}  (\pa_{x_1} \tilde\pa_{x_2})_{\a}{}^{\b} \, I(x_1,x_2,p)|_{p^2 = 0}\,.
\end{align}
 Introducing Schwinger parameters we obtain  
\begin{align}
I(x_1,x_2,p) = - \int^1_0 d\xi \int^\infty_0 d\eta\, \eta^{-1+\ep}\,  
e^{-\frac{p^2}{4\eta} - \xi \bar\xi \eta x_{12}^2 + i (p x_{1})\xi + i (p x_{2})\bar\xi}\ 
\end{align}
(cf. eq.~\p{2.4}). Substitution in \p{hlim} yields the anomaly 
\begin{align}
A_{\a}{}^{\b}(x_1,x_2,p) = 2\int^1_0 d\xi \, \xi\bar\xi \left[ 6 \delta_{\a}{}^{\b} + i \left(\xi (p\tilde x_{12})_{\a}{}^{\b} - \bar\xi (x_{12}\tilde p)_{\a}{}^{\b}) \right) \right] e^{i (p x_{1})\xi + i (p x_{2})\bar\xi} \,.
\end{align}

Now we transform the above results to momentum space, in order to clarify the collinear nature of the anomaly. The form factor \p{FYuk}
is given by the matrix product of two momentum space fermionic propagators
\begin{align}\label{}
\tilde F{}_{\a}{}^{\b}(q_1,q_2,p) = \frac{(q_1)_{\a\da}}{q^2_1} \frac{(\tilde q_2)^{\da\b}}{q^2_2} \ \delta(P)  \,,
\end{align}
where $P = q_1 +q_2 + p$ is the total momentum. The anomaly in the conformal Ward identity \p{WardYuk} takes the form of a one-parameter integral  
\begin{align}\label{326}
 2 i \pi^2 p_{\mu} \delta^{(4)}(P) \int^1_0 d\xi \,  \xi\bar\xi \left[ 6 \delta_{\a}{}^{\b} 
+ \xi  (p\, \tilde\pa_{q_1})_{\a}{}^{\b}  - \bar\xi (\pa_{q_1}\tilde p)_{\a}{}^{\b} \right]  \delta^{(4)}(q_1 + \xi p)  \,.
\end{align}
In close analogy with the 6D trivalent vertex \p{115}, the anomaly is supported on the collinear configuration $q_1 + \xi p=q_2+\bar\xi p =0$. The new element is the Lorentz tensor structure carried by the derivatives of the delta function.

\subsection{Gauge theory with scalar matter}\label{s33}

\begin{figure} 
\includegraphics[width = \textwidth]{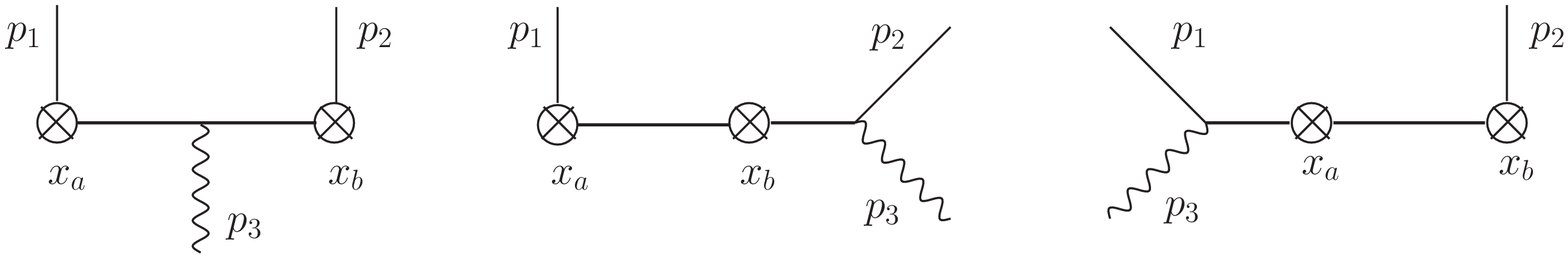}
\caption{Feynman diagrams contributing to the generalized form factor $F(x_a,x_b|p_1,p_2,p_3)$, eq.~\p{genFF3}, in the theory with the Lagrangian \p{Lagr}. }  \label{FigYMFF}
\end{figure}

In this subsection we present another example of an anomalous conformal Ward identity  in 4D conformal theories. We consider a massless scalar field in the adjoint representation of some gauge group,  interacting with the gauge field,
\begin{align}\label{Lagr}
L =\tr\left[ \frac12 (D_\mu \phi)^2 - \frac14 (F_{\mu\nu})^2\right]\,.
\end{align}
This theory is conformal at tree level. 
We define the gauge invariant operator $\cO= \tr \phi^2(x)$. Then we consider the Born level form factor of two such operators with an external state made of two scalars and one positive helicity gluon, 
\begin{align}\label{genFF3}
F(x_a,x_b|p_1,p_2,p_3):=\vev{\cO(x_a) \cO(x_b) |\phi(p_1) \phi(p_2) g^{(+)}(p_3) }_{\rm Born}\,.
\end{align} 
The computation is most easily done  in momentum space, i.e. after Fourier transforming  the two operators $\cO$ from position to momentum space, $x_{a,b} \to q_{a,b}$. The two types of tree-level Feynman diagrams are shown in Fig.~\ref{FigYMFF}.  
The result  is (up to a color factor)
\begin{align}\label{33}
\tilde F(q_a,q_b|p_1,p_2,p_3) = \delta^{(4)}(P)\frac{\vev{1|q_{a,1,3}\tilde q_{a,1}|2}}{\vev{13}\vev{23} q_{a,1}^2 q_{a,1,3}^2} +   (a\leftrightarrow b)\,,
\end{align}
where we use the shorthand notations $P=q_a+q_b+p_1+p_2+p_3$, $q_{a,i,\ldots,j} = q_a + p_i+ \ldots + p_j$, etc. This expression involves the off-shell momenta $q_a, q_b$ as well as the on-shell ones $p_i=\ket{i}[i|$. 
We remark that the Born-level generalized form factor of two weight-two half-BPS operators $\cO_{\bf 20'}$ in $\cN = 4$ SYM is given by the same expression.

In the form factor \p{33} we observe the familiar two-particle collinear poles of the type $\vev{ij}^{-1}$. As we have mentioned in the Introduction, they give rise to the so-called  ``holomorphic anomaly" in amplitudes \cite{Cachazo:2004by}.  Here we discuss the more general situation of a form factor, where some `legs' are off shell, the others are on shell. We wish to show that the collinear configurations of one off-shell and two on-shell legs cause new singularities which lead to  specific conformal anomaly terms. Unlike the obvious poles $\vev{ij}^{-1}$, the new singularities are hard to detect in momentum space. This is why we now Fourier transform the form factor \p{33}  back to coordinate space,  $ q_{a,b}  \to x_{a,b}$. 
The Fourier transform can be easily implemented by means of the formula\footnote{We thank Grisha Korchemsky for help with this integral.}
\begin{align} \label{Fourier1}
\int \frac{d^4 q}{4 \pi^2} \frac{e^{- i q x}\  \bra{\ell}q|p]  }{q^2 (q+p)^2} = \frac{e^{i x p} - 1}{2 i(x p)} \frac{\bra{\ell}x |p]}{x^2}\,.
\end{align}
The result is
\begin{align}\label{93}
F(x_a,x_b|p_1,p_2,p_3) = \frac{e^{i p_1 x_a + i p_2 x_b}}{(p_3 x_{ab})\,  x_{ab}^2} \left\{ \frac{ \bra{2}x_{ab} |3]}{\vev{23}} e^{i p_3 x_b}  - 
\frac{\bra{1}x_{ab} |3]}{\vev{13}} e^{i p_3 x_a} \right\}  + (a\leftrightarrow b)\ .
\end{align}
Note that the pole at $(p_3 x_{ab}) = 0$ is fake (its residue is zero), as can be seen by  expanding the exponentials. This is logical because the mixed $x/p$-space singularities are not physical. 

The form factor \p{93} is not invariant under conformal boosts with generator whose position space part $K^{(x)}_{\mu;\Delta=2}$ is defined in \p{a2} and the momentum part $\mK^{(p)}_\mu$ in \p{a9}.  In App.~\ref{s321} we derive the anomalous Ward identity\footnote{We ignore the holomorphic anomaly due to the poles $\vev{13} \vev{23}$, etc. in this calculation.}
\begin{align}\label{55}
\left( \sum_{i=a,b} K^{(x_i)}_{\mu;\Delta=2} + \sum_{i=1}^3 \mK^{(p_i)}_\mu  \right) F= - \frac{i}{4} e^{i p_1 x_a + i p_2 x_b} [3|\tilde\sigma_{\mu} x_{ab}|3]\,  A (x_a,x_b,p_3)  +  + (a\leftrightarrow b)\,,
\end{align}
where  
\begin{align} \label{h4D}
A (x_a,x_b,p_3)  := - \int^1_0 d \xi\, \xi \bar \xi\, e^{i \xi (x_a p_3) + i \bar \xi (x_b p_3)} \,.
\end{align}
A direct numerical calculation also confirms this result. Curiously, the expression for this 4D conformal anomaly  function coincides with the 6D one in  \p{27}.

The anomaly  is easier to obtain in the mixed $x/p$ representation  but its true nature is revealed in momentum space. As in the previous examples, the origin of the anomaly  is a hidden collinear singularity in the momentum space expression  \p{33}. To see it, we Fourier transform the anomaly term \p{55} back to  momentum space, $x_{a,b} \to q_{a,b}$, and find  
\begin{align}\label{}
&\left( \sum_{i=a,b} K^{(q_i)}_{\mu;\Delta=2}  +  \sum_{i=1}^3 \mK^{(p_i)}_\mu \right)  \tilde F  = \frac{i \pi^2}{2} \delta^{(4)}(P)[3|\tilde\sigma_\mu \pa_{q_a} |3]  \, \int_0^1 d\xi\, \xi \bar\xi\,   \delta^4(q_{a,1}+\xi p_3) +  (a\leftrightarrow b)\,. 
\end{align}
The anomaly has support on the configurations where the off-shell momenta become collinear with linear combinations of  two on-shell momenta, $q_a +p_1+\xi p_3 = q_b +p_2+\bar\xi p_3=0$.

\section{Conformal anomaly of loop integrals} 

An interesting corollary of the conformal anomaly of the Born-level vertices  is the conformal anomaly of certain Feynman integrals involving these vertices.
We consider several examples of one- and two-loop finite Feynman integrals, which are conformal if all the external legs are massive, $p^2_i\ne 0$.  We show that if some of the legs become massless, $p^2_i=0$, without causing IR divergences, the conformal symmetry is broken by the vertex anomaly. We  derive anomalous Ward identities in the form of 2nd-order inhomogeneous differential equations.  We check them against the known expressions for these integrals. In Sect.~\ref{s3} we consider the 6D one-loop boxes with different configurations of the external legs (on- or off-shell). In Sect.~\ref{s42} we study the 6D on-shell hexagon and in Sect.~\ref{secdblbox} the  4D double box with two on-shell and four off-shell legs. Finally, in App.~\ref{appD} we reinterpret the 6D conformal anomaly in terms of the discontinuities (cuts) of the integrals and of the `seed' vertex  \p{218}.

One-loop momentum integrals are conformal if  their legs are off shell, $p^2_i\neq0$. Indeed, consider the integral 
\begin{align}\label{41}
\cI_n(p_1,\ldots,p_n) = \delta^{(D)}(P)\,  \int \frac{d^D \ell}{\ell^2 (\ell + p_2)^2  (\ell + p_2 + p_3)^2 \ldots  (\ell + p_2+\ldots+p_{n})^2}\,,
\end{align}
where $P=\sum_{i=1}^n p_i$ is the total external momentum. If $n>D/2$ 
and all the external momenta are off shell, $p_i^2\neq0$, the integral is UV and IR finite. We can Fourier transform it to coordinate space. The result is a closed frame of free scalar propagators, 
\begin{align}\label{42}
\tilde \cI_n(x_1,\ldots,x_n) = \prod_{i=1}^n \frac1{(x^2_{i,i+1})^{D/2-1}}\,, \qquad x_{n+1} \equiv x_1\,.
\end{align} 
This expression is manifestly conformal with weight $\Delta_i = D-2$ at each point. Consequently, the function $I_n(p_1,\ldots,p_n)$ is also invariant under the action of the conformal boost generator \p{Kq}. The same is true for the reduced integral, where the delta function and $p_n$ are dropped,
\begin{align}\label{43}
I_n(p_1,\ldots,p_{n-1}) = \int \frac{d^D \ell}{\ell^2 (\ell + p_2)^2  (\ell + p_2 + p_3)^2 \ldots  (\ell + p_2+\ldots+p_{n-1})^2 (\ell - p_1)^2}\,.
\end{align}

It should be made clear that the conformal symmetry of these integrals has nothing to do with dual conformal invariance \cite{Broadhurst:1993ib,Drummond:2006rz}. The latter is a hidden symmetry of some loop integrals, while the former is just the native conformal symmetry of the theory. 

The question now is what happens if some legs are put on shell, $p^2_i=0$, for a subset $\{i\} \subset \{1,\ldots,n\}$. Depending on the space-time dimension, the integral may develop an IR singularity or remain finite. In the first case the regularization inevitably breaks conformal invariance, but what happens in the second case? This is what we wish to investigate here.

\subsection{Conformal anomaly of the 6D one-loop boxes}\label{s3}

In this and the next subsection we consider six dimensions. The one-loop integrals \p{43}  are finite for $D=6$, even with some or all of the external legs on the massless shell. Here we show that the  conformal symmetry of the off-shell integral  is in general spoiled by the collinear anomaly of the trivalent vertices with one massless leg.

\begin{figure} 
\begin{center}
\begin{tabular}{cccccc}
\includegraphics[width = 2.5cm]{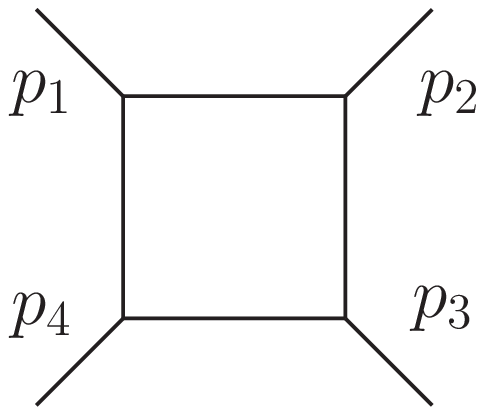} & \includegraphics[width = 2.5cm]{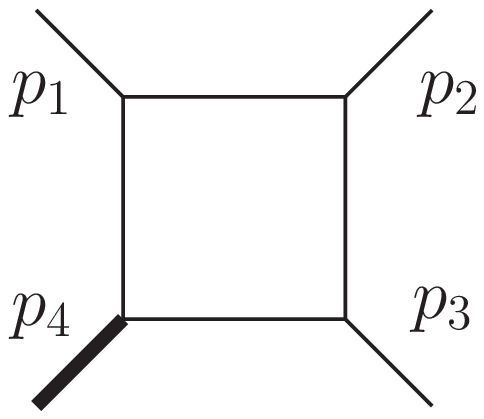} & \includegraphics[width = 2.5cm]{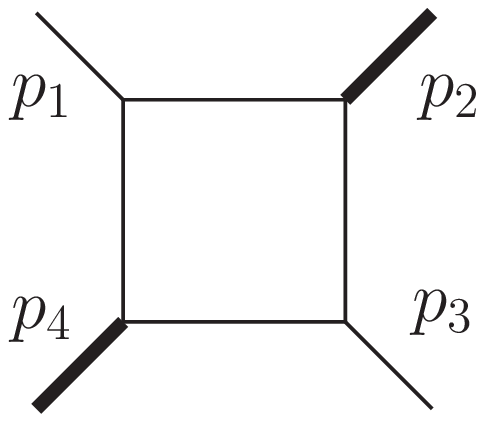} &
\includegraphics[width = 2.5cm]{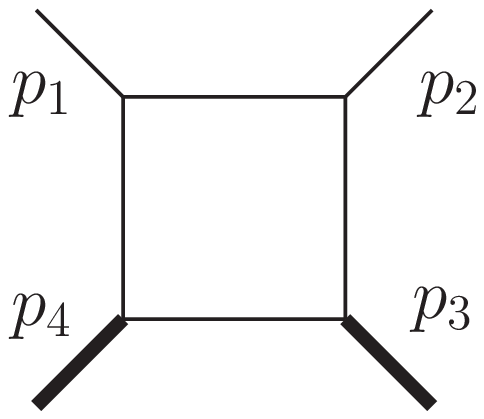} & \includegraphics[width = 2.5cm]{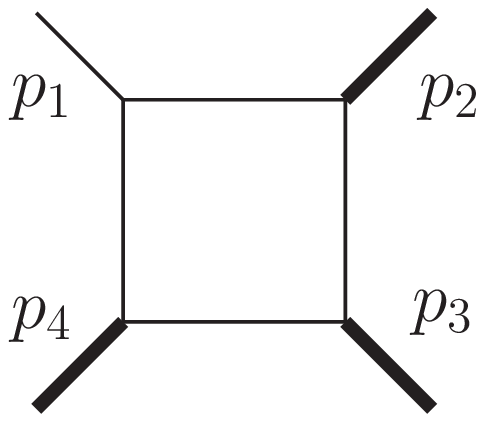} & \includegraphics[width = 2.5cm]{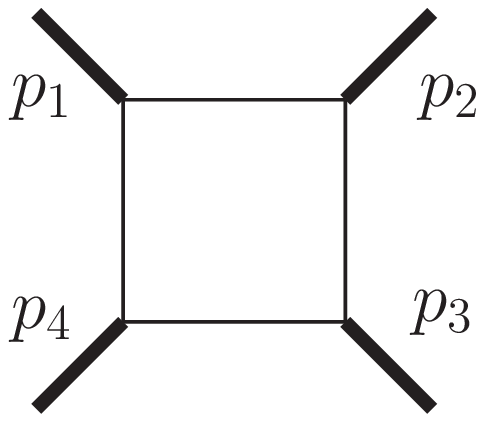} \\
$I_{0m}$ & $I_{1m}$ & $I_{2me}$ & $I_{2mh}$ & $I_{3m}$ & $I_{4m}$
\end{tabular}
\end{center}
\caption{6D box integrals. Thick external legs are off-shell and thin are on-shell momenta.}
\label{FigBoxInt}
\end{figure}

Consider first the one-loop  integral \p{43} of the box type, i.e. for $n=4$,
\begin{align}
I_{\text{box}}(p_1,p_2,p_3) = \frac{1}{i\pi^3}\int \frac{d^6 \ell}{\ell^2 (\ell + p_2)^2  (\ell + p_2 + p_3)^2 (\ell - p_1)^2}\,.
\end{align}
According to the configuration (on-shell/off-shell) of the external legs $p_1,\ldots,p_4$ one speaks of zero-mass, one-mass, two-mass-easy, two-mass-hard, three-mass, and four-mass box integrals, see Fig.~\ref{FigBoxInt}. Let us stress once more that unlike their 4D cousins, the 6D boxes are finite, because the trivalent vertices with a massless leg at the corners do not cause divergences in 6D. 

The box integrals admit the following Feynman parameter representation 
\begin{align} \label{alpha}
I_{\text{box}} = -\int_0^1 \prod_{l = 1}^4 d \alpha_l \frac{\delta\left( \sum_{k = 1}^4  \alpha_k -1 \right)}{\sum_{i<j} \alpha_i \alpha_j\,  y_{ij}^2}\,,
\end{align}
where the region momenta are defined by $y_i - y_{i+1} = p_i$, $y_5 \equiv y_1$, so $y_{ij}=p_{i,i+1,\ldots,j-1}$.
The iterated integrations over $\alpha$ can be performed resulting in dilogs and logs. However, the expressions are complicated and it is not obvious how to get a compact result. 

According to \cite{Bern:1992em} \footnote{We thank Zvi Bern for  the reference.} the 6D box integrals can be expressed in terms of the 4D box and triangle integrals. We find particularly simple 
formulae in the cases of the zero-mass, one-mass, and two-mass-easy boxes. In these cases the 4D triangles 
are absent and the 6D boxes are given by the finite part of the 4D boxes:
\begin{align}\label{3.2}
&I_{0m} = \frac{1}{2 p_{13}^2} \Bigl[ \log^2\left( \frac{p_{12}^2}{p_{23}^2} \right) + \pi^2 \Bigr]\,, \\
\label{3.3}
& I_{1m} = \frac{1}{p_{13}^2} \Bigl[ {\rm Li}_2\left(1-\frac{p_4^2}{p_{12}^2}\right) + {\rm Li}_2\left(1-\frac{p_4^2}{p_{23}^2}\right) 
+ \frac12\log^2\left(\frac{p_{12}^2}{p_{23}^2}\right) + \frac{\pi^2}{6} \Bigr]\,, \\
\label{34}
&I_{2me} = \frac{1}{p_{13}^2} 
\Bigl[ {\rm Li}_2\left(1-a p_2^2\right) + {\rm Li}_2\left(1-a p_4^2\right) - {\rm Li}_2\left(1-a p_{12}^2\right) - 
{\rm Li}_2\left(1-a p_{23}^2\right) \Bigr]\,,
\end{align}
where $a := \frac{p_2^2 + p_4^2 - p_{12}^2 -p_{23}^2}{p_2^2 p_4^2 - p_{12}^2 p_{23}^2}$.
The results for $I_{0m}$ and $I_{1m}$ can also be found in \cite{Anastasiou:1999cx}.\footnote{We thank Claude Duhr and Dmitri Kazakov for discussions of these integrals.}
The remaining 6D box integrals are more complicated   because of  the 4D triangle contributions.
The corresponding  formulae  can be extracted from \cite{Bern:1992em}.

We wish to show that the box integrals, like the three-point vertex \p{114}, satisfy anomalous conformal Ward identities with anomaly $A^{\mu}_{\text{box}}$,
\begin{align}\label{}
\left( \sum_{i=1}^4 K_i^\mu \right) \delta^{(6)}(P)  \, I_{\text{box}} = \delta^{(6)}(P)\, A^{\mu}_{\text{box}}\,,
\end{align}
where $K_{i}^{\mu}$ denotes  either $K^{\mu\; (p_i)}_{\Delta= 4}$ \p{Kq} for an off-shell leg or $\mK^{\mu\;(p_i)}$ \p{a12} for an on-shell leg. The conformal weight of the massive legs is $\Delta =6-2=4$, as explained after eq.~\p{42}. To put the identity in a more practical form we use the property \p{A1} and drop the delta function together with  one of the external momenta, $p_{i_4}$, $i_4 \in \{1,2,3,4\}$. Denoting the indices of the remaining momenta by $i_1,\,i_2,\,i_3 \in \{1,2,3,4\} \backslash \{i_4\}$ we obtain
\begin{align}\label{3gen}
\left( K_{i_1}^\mu + K_{i_2}^\mu + K_{i_3}^\mu \right) \, I_{\text{box}}(p_{i_1},p_{i_2},p_{i_3}) 
=  A^{\mu}_{\text{box}}(p_{i_1},p_{i_2},p_{i_3})\,,
\end{align}
where $p_{i_4} = - p_{i_1} - p_{i_2} - p_{i_3}$.

\begin{figure}
\begin{align*}
\left( \sum_{i=1}^4 \mK^\mu_i \right) \begin{array}{c}\includegraphics[width = 3cm]{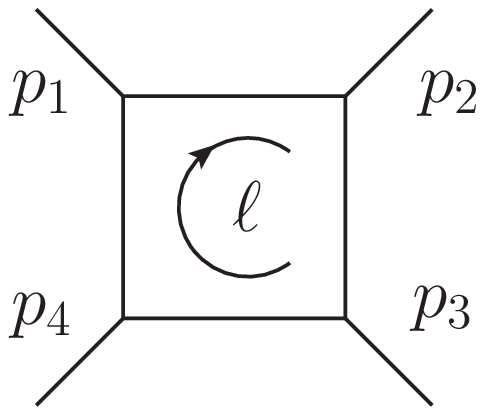}\end{array} & = \begin{array}{c}\includegraphics[width = 3cm]{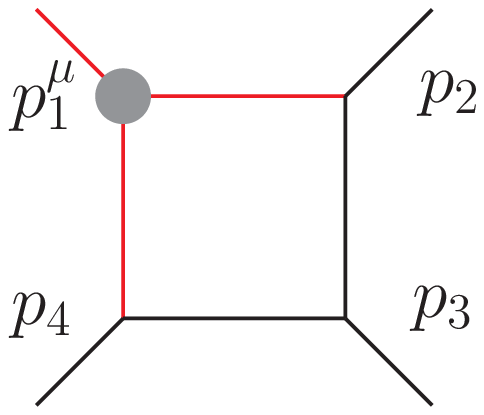}\end{array} + \begin{array}{c}\includegraphics[width = 3cm]{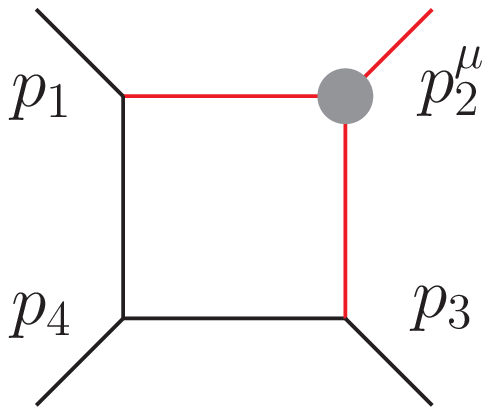}\end{array} \\[0.7cm]
& + \begin{array}{c}\includegraphics[width = 3cm]{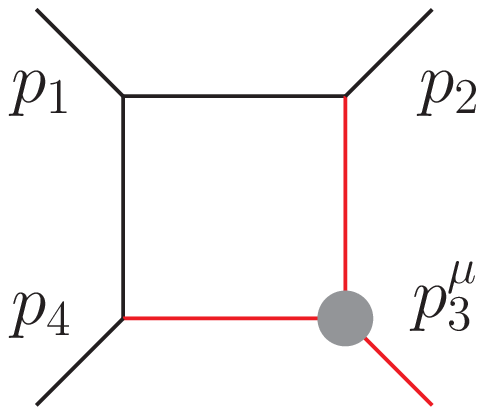}\end{array} + \begin{array}{c}\includegraphics[width = 3cm]{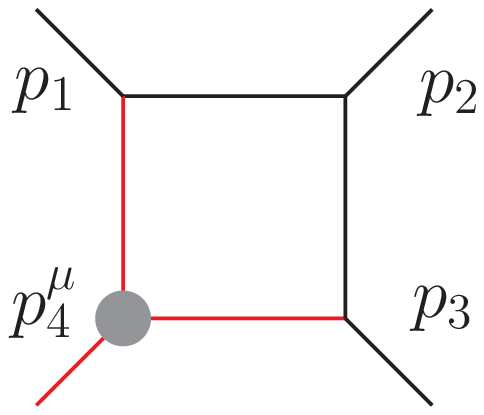}\end{array} = 0
\end{align*}
\caption{Cancellation of the conformal anomalies of the 6D zero-mass box. The momenta of the highlighted lines are collinear. }
 \label{fig0m}
\end{figure}
\begin{figure} 
\begin{align*}
\left( \mK^\mu_1 + K^\mu_{2} + \mK^\mu_3 + K^\mu_{4} \right) \begin{array}{c}\includegraphics[width = 3cm]{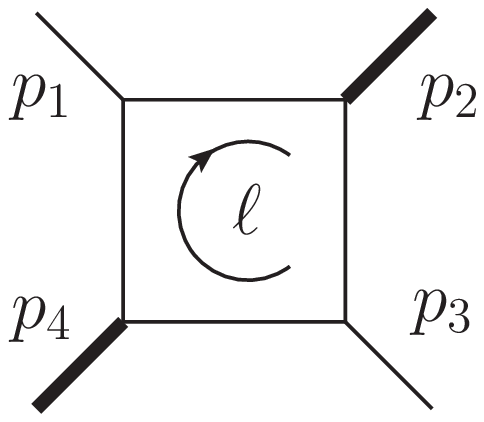}\end{array} & = \begin{array}{c}\includegraphics[width = 3cm]{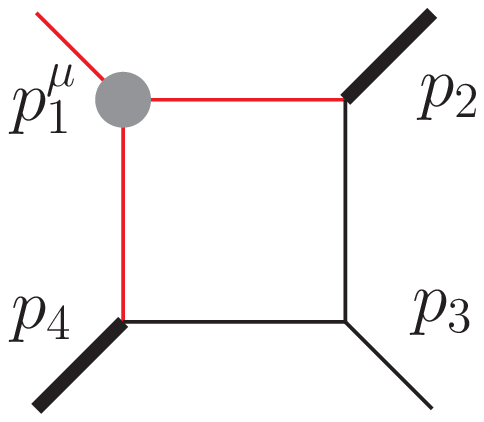}\end{array} + \begin{array}{c}\includegraphics[width = 3cm]{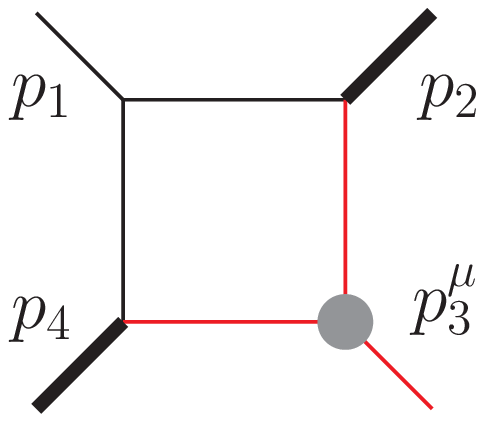}\end{array} \neq  0
\end{align*}
\caption{Anomalous conformal Ward identity for the two-mass-easy 6D box integral.}
\label{fig2me}
\end{figure}

The explicit form of the anomaly $A^{\mu}_{\text{box}}$ depends on the configuration of the external legs. The four-mass box is conformal since all legs are off-shell, i.e. $A^{\mu}_{4m}=0$. 
If one or more legs are on-shell, the integration over the regions where the loop momentum $\ell$ is collinear with the on-shell leg, in general spoils the  symmetry. The zero-mass box $I_{0m}$ \p{3.2} is an exception. In this particular case the contributions from the four collinear loop integration regions cancel in the sum and the integral is conformal, as explained below.

The conformal anomaly $A^{(6D)}_{\mu}(p;\ell)$ of the trivalent vertex at a massless corner with inflowing loop momentum $\ell$ and  on-shell momentum $p$  is given by \p{218}. We act on each  leg of $I_{0m}$ with the conformal boost. If we ignore the anomaly, the transformation of the external legs  is compensated by that of the loop momentum, since we know that the four-mass  box is conformal.   The effect of the anomaly amounts to replacing  each vertex by the contact term $A^{(6D)}_{\mu}$, see Fig.~\ref{fig0m}. This localizes the  loop integration on the configuration of  collinear momenta. The remaining integration over $\xi$ from \p{218} is straightforward,
\begin{align}
\mK_i^\mu \, I_{0m} \Rightarrow \int d^6 \ell\, \frac{A_{\mu}^{(6D)}(p_i;\ell)}{(\ell - p_{i+1})^2 (\ell - p_{i+1} - p_{i+2})^2} = \frac{4 i \pi^3 p^\mu_i}{p_{i-1 i}^2 p_{i i+1}^2}\,.
\end{align}
Summing up all four contributions, we find that the anomaly cancels, 
\begin{align}
\left( \sum_{i=1}^4 \mK_i^\mu \right) \delta^{(6)}(P) I_{0m} = 4\, \delta^{(6)}(P)\, \sum_{i=1}^4 \frac{p^\mu_i }{p_{i-1 i}^2 p_{i i+1}^2} = 0\,.
\end{align}
The direct application of the conformal boost $\mK^\mu$ from \p{a12} on the expression \p{3.2} confirms that the massless 6D box integral has no anomaly.

The conformal symmetry of the remaining 6D boxes in Fig.~\ref{FigBoxInt} with on-shell legs is broken by the collinear anomaly. 
The corresponding Ward identities are obtained as above. We act with  $\mK^\mu$ on the on-shell legs and with $K^\mu_{\Delta = 4}$  on the 
off-shell legs. Each massless corner $i$ is replaced by the anomaly \p{218} which freezes the loop momentum, see Fig.~\ref{fig2me}. The result of the  integration over $\xi$ depends on the configuration of the adjacent legs $i-1$, $i+1$:
\begin{align}
&A^{i\,\mu}_{\text{on/on}} = \frac{4 p^\mu_i}{p_{i-1 i}^2 p_{i i+1}^2} =: a^{\mu}\,,\nt
&A^{i\,\mu}_{\text{off/on}} = \frac{a^{\mu} }{1-r_-} \left( 1 + \frac{r_-\log r_-}{1-r_-}\right) 
\qquad \text{with} \qquad {r_- = \frac{p^2_{i-1}}{p^2_{i-1 i}}},\nt
&A^{i\,\mu}_{\text{on/off}} =  \frac{a^{\mu} }{1-r_+} \left( 1 + \frac{r_+\log r_+}{1-r_+}\right) 
\qquad \text{with} \qquad {r_+ = \frac{p^2_{i+1}}{p^2_{i i+1}}},\nt
&A^{i\,\mu}_{\text{off/off}} = \frac{a^\mu}{1-r_- r_+} \left( \frac{r_- \log r_-}{(1- {r_-})^2}  + \frac{r_+ \log r_+}{(1-{r_+})^2} \right) + \frac{a^\mu}{(1-r_+)(1-r_-)}\,. \label{boxAnom'}
\end{align}
Summing up the relevant anomaly terms, we find the following conformal Ward identities for the remaining box integrals, 
\begin{align}
&\left( \mK^\mu_1 + \mK^\mu_2 + \mK^\mu_3 + K^\mu_4\right) \delta^6(P)\, I_{1m} = \delta^6(P) \left( A^{1\, \mu}_{\text{off/on}} + A^{2\, \mu}_{\text{on/on}} + A^{3\, \mu}_{\text{on/off}} \right), \label{1mWI}\\
&\left( \mK^\mu_1 + K^\mu_2 + \mK^\mu_3 + K^\mu_4\right) \delta^6(P)\, I_{2me} = \delta^6(P) \left( A^{1\, \mu}_{\text{off/off}} + A^{3\, \mu}_{\text{off/off}} \right), \label{2meWI} \\
&\left( \mK^\mu_1 + \mK^\mu_2 + K^\mu_3 + K^\mu_4\right) \delta^6(P)\, I_{2mh} =\delta^6(P) \left( A^{1\, \mu}_{\text{off/on}} + A^{2\, \mu}_{\text{on/off}} \right), \\
&\left( \mK^\mu_1 + K^\mu_2 + K^\mu_3 + K^\mu_4\right) \delta^6(P)\, I_{3m} = \delta^6(P) \, A^{1\, \mu}_{\text{off/off}}\,.
\end{align}
We have checked these identities  explicitly, in the form \p{3gen} by acting with the conformal boost  directly on the expressions \p{3.3} and \p{34} for the one-mass and two-mass-easy boxes; for the remaining integrals we  performed the $\alpha$-parameter  integration \p{alpha} numerically.

In conclusion, the one-loop 6D boxes are  {not conformal}, except for the four-mass and zero-mass cases. 
The breakdown is due to the collinear anomaly that we have revealed. We can predict this anomaly without actually knowing the expression for the integral itself. 

It would be interesting to investigate if the anomalous conformal Ward identities can be turned into useful differential equations in momentum space. They will relate the $\ell-$loop (pseudo)conformal integrals to $(\ell-1)-$loop ones.  Solving such equations could be an alternative way of calculating  (pseudo)conformal loop integrals.

\subsection{Conformal anomaly of the 6D hexagon integral}\label{s42}

\begin{figure}
\begin{align*}
\left( \sum_{i=1}^6 \mK^\mu_i \right) \begin{array}{c}\includegraphics[height = 3.0cm]{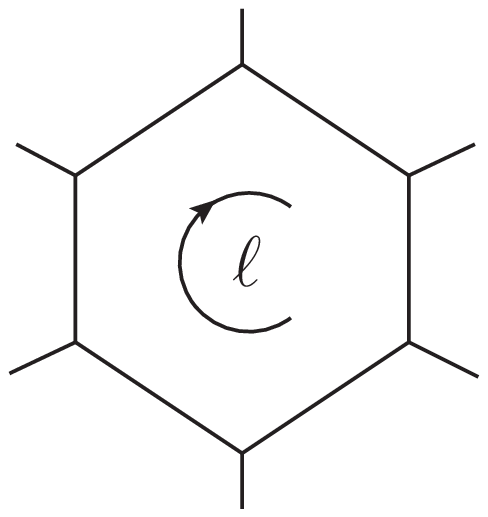}\end{array} \;\;\; = \;\;\; \sum_{i=1}^6 \; 
\begin{array}{c}\includegraphics[height = 3.0cm]{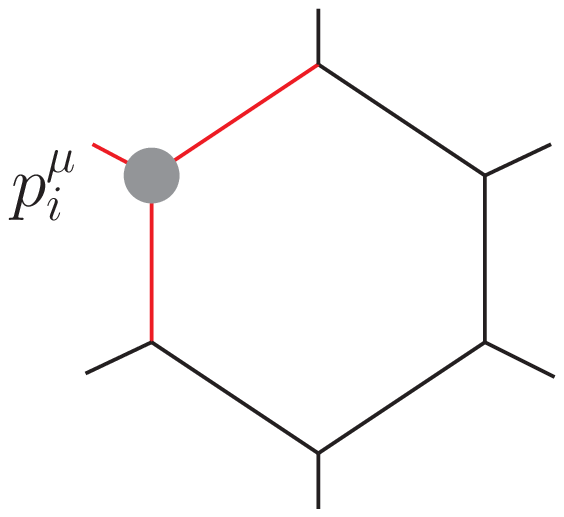}\end{array}
\end{align*}
\caption{Conformal Ward identity for the 6D hexagon integral with all legs on shell.} \label{FigHexWI}
\end{figure}

Another interesting example is the 6D on-shell hexagon integral, $p_i^2 = 0$, $i=1,\ldots,6$,  
\begin{align}
I_{\rm hex} = \frac{1}{i \pi^3}\int \frac{d^6 \ell}{\ell^2 (\ell + p_1)^2 (\ell + p_{1,2})^2 (\ell + p_{1,2,3})^2 (\ell - p_{5,6})^2 (\ell - p_{6})^2}\,.
\end{align}
This integral is  finite. If all the legs are massive, the integral is conformal, see \p{43}. In the massless case the collinear anomaly breaks the conformal symmetry in a predictable way. 

Besides the (anomalous) conformal symmetry, this integral is also dual conformal \cite{Broadhurst:1993ib}, \cite{Drummond:2006rz}. As a corollary, it is a function of three cross-ratios,  and is expressed in terms of weight three polylogarithms, see  \cite{Dixon:2011ng}, \cite{DelDuca:2011ne}.
Like in Sect.~\ref{s3}, using the seed anomaly \p{218} at each vertex, we obtain the conformal Ward identity  depicted schematically in Fig.~\ref{FigHexWI},\footnote{The possibility of a conformal anomaly in momentum space is mentioned at the end of Ref.~\cite{Dixon:2011ng}.} 
\begin{align}
\left( \sum_{i=1}^6 \mK^\mu_i \right) \delta^6(P)\, I_{\rm hex} =4 \,\delta^6(P) \sum_{i = 1}^6 \frac{p_i^\mu}{p_{i,i-1}^2 p_{i,i+1}^2} \frac{\log\left( \frac{p_{i+1,i+2}^2 p_{i-1,i-2}^2}{p_{i,i+1,i+2}^2 p_{i,i-1,i-2}^2}\right)}{(p_{i+1,i+2}^2 p_{i-1,i-2}^2 - p_{i,i+1,i+2}^2 p_{i,i-1,i-2}^2)}\,. \label{HexWI}
\end{align}
We have checked  it using the explicit functional expression for $I_{\rm hex}$ from \cite{Dixon:2011ng}.

\subsection{Conformal anomaly of the 4D six-leg double box integral}\label{secdblbox}

\begin{figure}
\begin{align*}
\left( \sum_{i=1}^6 K^\mu_i \right) \begin{array}{c}\includegraphics[width = 4cm]{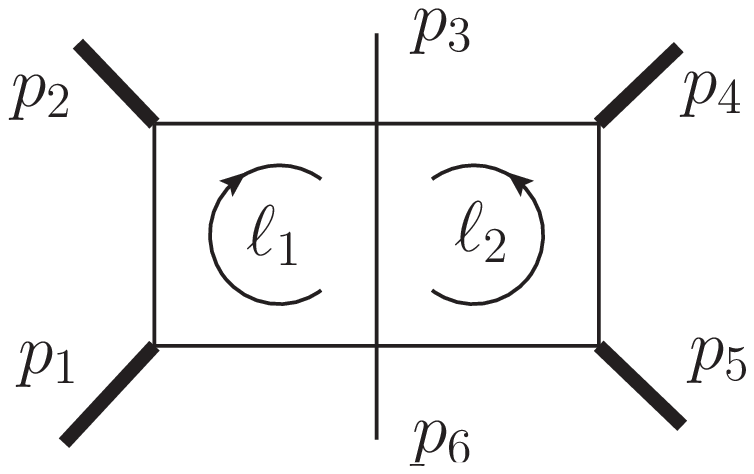}\end{array} = \begin{array}{c}\includegraphics[width = 4cm]{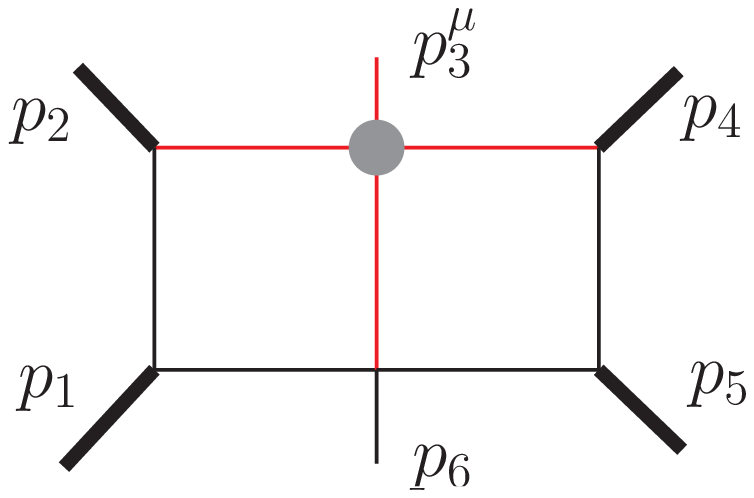}\end{array} + \begin{array}{c}\includegraphics[width = 4cm]{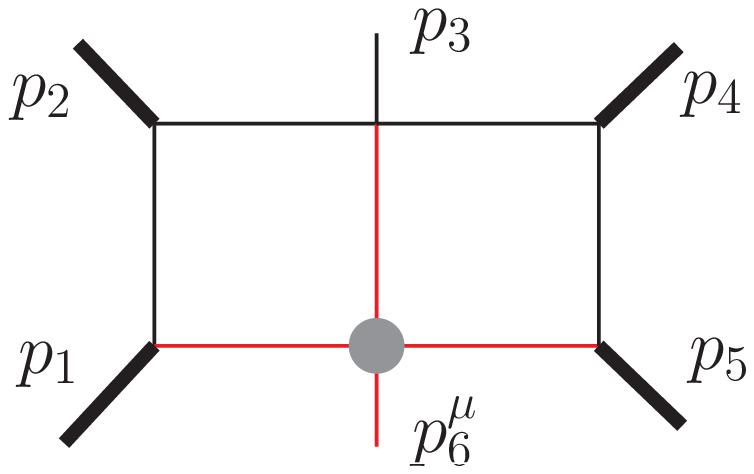}\end{array}
\end{align*}
\caption{Conformal Ward identity for the 4D double box. We act with $\mK^\mu$ on the on-shell legs $p_3$, $p_6$, and with  $K^\mu_{\Delta=2}$ on the off-shell legs $p_1$, $p_2$, $p_4$, $p_5$.}\label{f6}
\end{figure}

Another interesting example of a finite integral is the 4D double box depicted in Fig.~\ref{f6},
\begin{align}\label{420}
I^{\text{dbl}}_{\text{box}}(p_1,\ldots,p_6) = \frac{1}{4 \pi^4}\int \frac{d^4 \ell_1\, d^4 \ell_2}{\ell_1^2 \ell_2^2 (\ell_1-p_2)^2 (\ell_1-p_{1,2})^2 (\ell_2-p_4)^2 (\ell_2-p_{4,5})^2 (\ell_1+\ell_2+p_3)^2 }\,.
\end{align}
Unlike the previous examples, the analytic answer for this integral is unknown and it is believed not to be expressible in terms of harmonic polylogarithms. In particular, the maximal cut of the double box integral is given by an elliptic integral \cite{CaronHuot:2012ab}.

 If all of its legs are off shell, it is the Fourier transform of a frame of free scalars propagators $1/x^2_{ij}$, with the same topology and with conformal weights $\Delta=2$ at the corners and $\Delta=3$ at the middle points. This shows that the off-shell integral is conformal. 
 
 We are interested in the case where the two middle  legs  are massless, $p_3^2 =0$, $p_6^2 = 0$.\footnote{In the literature this integral is referred to as the ten-leg double box, implying that the massive as well as the massless legs come from $\phi^4$ vertices. We count six legs, four massive and two massless.} The integral remains IR finite but its conformal symmetry is broken. As shown in the figure, around the massless legs we find 4D `seed' configurations of four momenta which have a collinear anomaly, see \p{314}. Acting with the conformal boost $\sum_{i=1,2,4,5} K_{i\,; \Delta=2}+\mK_3 + \mK_6$, we produce contact terms which lift one of the loop integrations. The anomaly can be put in the following Feynman parameter form
\begin{align}
\sum_{i=3,6} p_i^\mu \int_0^1 \frac{d \alpha \, d \beta \, d \gamma \, \delta(\alpha+\beta+\gamma-1)}{(\alpha y_{i-1\,i+1}^2 + \bar\alpha y_{i-1\,i}^2)(\alpha y_{i-2\,i+1}^2 + \bar\alpha y_{i-2\,i}^2) (\beta y_{i\,i+2}^2 + \bar\beta y_{i+1\,i+2}^2) (\beta y_{i\,i+3}^2 + \bar\beta y_{i+1\,i+3}^2)}\,,
\end{align}
where the region momenta are defined by $y_i - y_{i+1} = p_i$, $y_7 \equiv y_1$, so $y_{ij}=p_{i,i+1,\ldots,j-1}$. { It is straightforward to reduce this integral to polylogs of weight two.}

We have checked this anomalous Ward identity by numerical integration.
We used a Feynman parameter representation for the double-box integral, similar to \p{alpha}, 
and acted on it with the  conformal boost generators. 
Then we chose random kinematics in the Euclidean region and performed the integrations.

In conclusion, our method produces a potentially useful 2nd-order differential equation for the double box integral, whose right-hand side is given by polylogs of weight two. {It is interesting to clarify the relationship with the 1st-order differential equation from \cite{Nandan:2013ip}.}

\section{Conclusions}

In this paper we have revealed a new mechanism of breakdown of conformal invariance, at the lowest, Born level of perturbation theory, hence in the absence of UV or IR/collinear divergences. The phenomenon occurs in generalized form factors involving more than one local operator and an on-shell state of massless particles. The breakdown is due to hidden singularities on configurations  in momentum space where the momenta of the operators become collinear with the on-shell momenta of the particles. The contact nature of the conformal anomaly makes it difficult to detect directly in momentum space. It is much easier to see in the mixed  representation, where the operators live in coordinate and the particles in momentum space. There the anomaly is not of the contact type and can be most efficiently worked out by the method of Lagrangian insertion. We have presented a number of examples in 4D and 6D conformal theories. 

We have found a practical application of the new conformal anomaly to the study of loop momentum integrals. It concerns a class of $\ell-$loop integrals which remain finite if some of their legs become massless. The integrals are conformal if their legs are massive but the collinear region  around a massless external leg creates a contact anomaly of the type we are discussing. The loop integration makes this anomaly visible. It takes the form of a 2nd-order differential equation whose anomalous right-hand side is given by $(\ell-1)-$loop integrals. We have verified this effect on several examples of one- and two-loop scalar integrals. Our differential equations might prove useful for calculating unknown finite loop integrals. Differential equations of a different origin have been successfully exploited in Refs.~\cite{Drummond:2006rz}, \cite{Drummond:2010cz}, \cite{Dixon:2011ng}. It would be very interesting to combine the two methods.

In this paper we have restricted ourselves to 4D and 6D conformal theories but it is straightforward to extend our results to 3D form factors based on the conformal $\phi^6$ vertex. Another line of generalization is to study not only scalar finite integrals but also those with fermion lines. We can expect similar anomalies and differential equations. 

Our discussion here concerns only ordinary conformal symmetry. It can be   extended to the maximally supersymmetric $\cN = 4$ SYM theory, which is conformal at all levels of perturbation theory. {In this theory there exist Born-level generalized form factors with loop topology, so their conformal anomaly is expected to be non-contact.} One should also encounter an anomaly of the superconformal symmetry ($S$ or $\bar S$).  It would be interesting to clarify the relation to the dual $\bar Q$, or equivalently, $S$ supersymmetry of amplitudes discussed in  Ref.~\cite{CaronHuot:2011kk}.  Conformal and dual conformal (super)symmetry are at the base of the Yangian symmetry of superamplitudes \cite{Drummond:2008vq,Drummond:2009fd} and more recently, of the multi-loop fishnet graphs \cite{Chicherin:2017frs}. Our new anomaly mechanism may have implications for these larger symmetries too. {In particular, the considerations in the present paper should be sufficient to describe the anomalies of the level one Yangian symmetry generators of the fishnet graphs}. 

\section*{Acknowledgements}

We profited from numerous discussions with Zvi Bern,  Simon Caron-Huot, Claude Duhr, Dmitri Kazakov, Grisha Korchemsky, Radu Roiban and Ivan Todorov. E.S. is  grateful to the School of Physics and Astronomy at Queen Mary University of London for hospitality, and in particular to Andy Brandhuber and Gabrielle Travaglini, whose questions triggered this investigation. The  work of D.C. has been   partially supported by the Russian Science Foundation, grant N 14-11-00598.

%\newpage

\section*{Appendices} 

\appendix

\section{Conformal generators in coordinate and momentum space}\label{appA}

In this paper we consider the $D-$dimensional  conformal group $SO(2,D)$ realized  in coordinate and momentum space. We focus particularly on the cases $D=4$ and $D=6$.

The familiar coordinate space realization of the translation $P_\mu$ and conformal boost $K_\mu$ generators on a function of several points $\varphi(x_i)$ with conformal weight $\Delta_i$ and $D-$dimensional Lorentz spin  $S_i$  at  each point,  has the form
\begin{align}\label{a1}
&  P^{(x)}_{\mu} = i\sum_i\pa_{x^\mu_i} \\
& K^{(x)}_{\mu;\Delta,S} =  i\sum_i \left(x^2_i \pa_{x^\mu_i} - 2 x_{i\, \mu} x^{\nu}_i \pa_{x^\nu_i} - 2 \Delta_i x_{i\, \mu} -2i x^\nu_i \Sigma_{i\, \nu\mu}\right) \,. \label{a2}
\end{align}
Here $\Sigma_{i}$ is the matrix part of the Lorentz generator $L_{\mu\nu} $ corresponding to the given representation.   They satisfy the conformal algebra  $[P_\mu, K_\nu]= 2iL_{\mu\nu} + 2i\eta_{\mu\nu} {\cal D}$, where the dilatation operator is ${\cal D} = -i\sum_i \left( x^{\mu}_i \pa_{x^\mu_i} +  \Delta_i  \right) $. The generators are defined so that the propagator of a massless scalar field with canonical conformal weight $\Delta = D/2-1$,  $\vev{\phi(x_1)\phi(x_2)} =  {(x^2_{12})^{1-D/2}} $, is invariant,
\begin{align}\label{}
\{P,K,L,{\cal D} \}\, \vev{\phi(x_1)\phi(x_2)} =  0\,. 
\end{align}

The momentum space realization is obtained by Fourier transforming the $x$-space conformal generators \p{a1}, \p{a2} according to the following rule with a test function $\varphi(x)$, 
\begin{align}
G^{(q)} \,\tilde \varphi(q) := \int d^D x \, e^{i q x} G^{(x)} \,\varphi(x) \qquad \text{where} \quad
\tilde \varphi(q) := \int d^D x \,e^{i q x} \varphi(x)\,.
\end{align}
We find 
\begin{align} \label{Pq}
&  P^{(q)}_{\mu} = \sum_i q_{i\, \mu}\\
& K^{(q)}_{\mu;\Delta} = \sum_i  \left[ -q_{i\, \mu} \Box_{q_i} + 2 q^\nu_i \pa_{q^\nu_i} \pa_{q^\mu_{i}} + 2(D-\Delta_i)\pa_{q^\mu_i} + 2i \Sigma_{i,\mu\nu} \pa_{q_{i\,\nu}}\right].  \label{Kq}
\end{align}

The Fourier exponential $e^{iqx}$ is invariant under both generators,
\begin{align}\label{}
(P^{(q)}_\mu + P^{(x)}_\mu ) \, e^{iqx}=0\,, \qquad  (K^{(q)}_{\mu; \Delta} + K^{(x)}_{\mu; D-\Delta} ) \, e^{iqx}=0\,.
\end{align}
The arbitrary choice of the conformal weight $\Delta$ reflects the property of $\delta^{(D)}(x_{12})$ (the inverse Fourier transform of $e^{iqx_{12}}$). Its total weight is $D$, but the individual weights at points 1 and 2 cannot be distinguished. 

Since the conformal boost generator \p{Kq} is a 2nd-order operator, the product of two conformal functions of the momenta is  in general not conformal. A useful illustration is the product  of a scalar propagator $1/{q^2}$ and some other scalar function $\varphi(q)$. We have
\begin{align}\label{a8}
K^{(q)}_{\mu;\Delta}\, \left[\frac1{q^2}  \varphi(q)\right] = & 4 \left(\Delta+1-\frac{D}{2}\right) \frac{q_\mu}{q^4}\varphi(q)\nt
&+ \frac{1}{q^2}\left[-q_\mu \Box +2q^\nu \pa_\nu \pa_\mu + 2(D-\Delta-2)\pa_\mu \right] \varphi(q)\,.
\end{align}
We see that this product is conformal only if $\Delta =D/2-1$, i.e. the conformal weight of a scalar field in $D$ dimensions. Consequently, the function $ \varphi(q)$ must have weight $\Delta_\varphi = D/2+1$. This is the momentum space equivalent of the coordinate space statement that the equation $\Box_x \phi(x) = \varphi(x)$ is conformal only if $\Delta_\phi =D/2-1$ and $\Delta_\varphi = D/2+1$.

\subsection{On-shell realization}

The conformal group can also be realized  on functions of on-shell momenta $\varphi(p)$ with $p^2=0$. This realization depends on the space-time dimension. Here we give the details for the cases of interest in this paper,  $D=4$ and $D=6$.

A real lightlike momentum $p_\mu$, $p^2 = 0$ in 4D Minkowski space is parametrized 
by a complex conjugate pair of commuting  chiral and antichiral $SL(2,C)$ spinors, $\sigma^{\mu}_{\a\da} p_{\mu} = \la_{\a} \tilde\la_{\da}$. They are defined up to a $U(1)\sim SO(2)$ phase which is associated with the helicity of the on-shell particle. This $SO(2)$ is the little group (subgroup of $SL(2,C)$) which leaves $p_\mu$ invariant.  In this parametrization the on-shell conformal  generators take the form \cite{Witten:2003nn} 
\begin{align} \label{a9}
\mathbb P_{\mu} = \sum_i p_{i\, \mu} =  \frac12 \, \tilde\sigma^{\da\a}_{\mu} \sum_i \la_{i\, \a} \tilde\la_{\, \da} \,, \qquad  \mK_{\mu} = 2 \, \tilde\sigma^{\da\a}_{\mu} \sum_i \frac{\pa^2}{\pa \la^{\a}_i \pa \tilde\la^{\da}_i}\,.
\end{align}
A short calculation using the chain rule shows that
\begin{align}\label{a10}
\mK_{\mu} \, \varphi(p) = K^{(p)}_{\mu; \Delta = 3} \, \varphi(p) \ \ {\rm for} \ D=4\,,
\end{align}
on the space of on-shell test functions $\varphi(p_{\a\da} = \la_{\a}\tilde\la_{\da})$.
Thus the off-shell, eq.~\p{Kq}, and on-shell, eq.~\p{a9}, versions of the conformal boost are compatible.

In the case $D=6$ we use the commuting spinor parametrization of a 6D lightlike vector of Ref.~\cite{Cheung:2009dc}. 
The complexification of the 6D Lorentz group is $SL(4)$, and the corresponding little group of a lightlike vector is $SL(2) \times SL(2)$. We use chiral spinors $\lambda^{Aa}$ carrying an $SL(4)$ index $A$ and an $SL(2)$ index $a$ of the little group labeling the helicity states. The vector  representation of $SO(6)$ is equivalent to the antisymmetric bispinor representation  of $SL(4)$, so $p_\mu$ is given by the
antisymmetric product of two chiral spinors, $p^{\mu}\tilde \sigma_{\mu}^{AB} = \lambda^{Aa} \lambda^{B}_{a}$. The Clebsch-Gordon coefficients $\tilde\sigma_{\mu}^{AB}$ are the $4 \times 4$ 6D Pauli matrices (see App.~A in \cite{Cheung:2009dc}). In this spinor parametrization the on-shell conformal boost generator takes the form 
\begin{align} \label{a11}
\mK_\mu = - \tilde\sigma_\mu^{AB} \frac{\pa^2}{\pa \lambda^{Aa} \pa \lambda^{B}_{a}}\,.
\end{align}
On the space of on-shell test functions $ \varphi\left(p^{AB} = \lambda^{Aa} \lambda^{B}_{a}\right)$ this is equivalent to 
\begin{align} \label{a12}
\mK_{\mu}\, \varphi(p) =K^{(p)}_{\mu;\Delta=4} \,\varphi(p)  \ \ {\rm for} \ D=6\,.
\end{align}

\subsection{Conformal properties of the momentum conservation delta function}\label{appA2}

Here we prove that the momentum conservation delta function in expressions of the type $\delta^{(D)}(\sum_{i=1}^n q_i) \varphi(q_i)$ can be dropped when checking the conformal properties. It is sufficient to show that $K^{(q)}_\mu \varphi(q)=0$, where $\varphi(q)$ depends on $(n-1)$ momenta. For simplicity, we consider the case $n=2$, the generalization is straightforward.

We want to show that
\begin{align}\label{A1}
(K^{(q)}_\mu + K^{(k)}_\mu)\, \left[\delta^{(D)}(q+k)\, \varphi(k)  \right] = 0 \ \ {\rm iff} \ \ K^{(k)}_\mu\, \varphi(k)=0\,.
\end{align}
Here we also assume that $\varphi(k)$ is Lorentz invariant and homogeneous of degree $\Delta_k$,
\begin{align}\label{A2}
L_{\mu\nu}\varphi \sim (k_\mu \pa_{k^\nu} - k_\nu \pa_{k^\mu})\varphi=0\,, \qquad   k^\nu \pa_{k^\nu}\varphi = \Delta_k \varphi\,.
\end{align} 
The conformal generator in $D-$dimensional momentum space is given in \p{Kq}, with $\Delta_i$ being the conformal weight associated with the point $x_i$ whose Fourier dual is $q_i$. The operator $K^{(q)}_\mu $ in \p{A1} acts only on the delta function whose conformal weight is $\Delta_q=D$. Switching the derivatives from its $q$ end to the $k$ end and integrating by parts, we find
\begin{align}\label{}
K^{(q)}_\mu   \, \delta^{(D)}(q+k) = \left[k_\mu \Box_k - 2 k^\nu \pa_{k^\nu} \pa_{k^\mu}   -2D  \pa_{k^\mu} \right]   \, \delta^{(D)}(q+k) \,,
\end{align}
therefore
\begin{align}\label{A4}
(K^{(q)}_\mu + K^{(k)}_\mu)  \, \delta^{(D)}(q+k) =   -2 \Delta_k  \pa_{k^\mu}  \, \delta^{(D)}(q+k) \,.  
\end{align}

The operator $K^{(k)}_\mu $ acts also on the function $\varphi(k)$. We assume that it is invariant, i.e. for some $\Delta_k$ we have $K^{(k)}_\mu\, \varphi(k)=0$. What remains are the mixed terms where the second-order derivatives in $K^{(k)}_\mu $ are distributed between $\delta^{(D)}(q+k)$ and $\varphi(k)$:
\begin{align}\label{A5}
-k_\mu \Box_k \, (\delta \varphi) \ &\rightarrow \ -2k_\mu (\pa_{k_\nu}\delta)\, (\pa_{k^\nu} \varphi) \nt
  2 k^\nu \pa_{k^\nu} \pa_{k^\mu}\, (\delta \varphi) \ &\rightarrow \ 2 k^\nu( (\pa_{k^\mu}\delta)\, (\pa_{k^\nu}\varphi) + (\pa_{k^\nu}\delta)\, (\pa_{k^\mu}\varphi) ) \nt &= 2(\Delta_k\, (\pa_{k^\mu}\delta)\,  \varphi + k_\mu (\pa_{k^\nu}\delta)\, (\pa_{k_\nu}\varphi) )\,.
\end{align}
Going from the second to the third line we have used the properties \p{A2} of the function $\varphi(k)$. So, the net result from \p{A5} is $2\Delta_k\, (\pa_{k^\mu}\delta)\,  \varphi$, which cancels the delta function contribution    \p{A4}, and we arrive at \p{A1}. 

The same argument works if the momentum $k_\mu$ is on-shell, $k^2=0$. After distributing the derivatives  from the generator $ K^{(k)}$ \p{a10} or \p{a12}, we use the analogs of \p{A2}.

\section{Calculation of the 6D $\phi^3$ form factor and its anomaly}\label{appB}

Here we present a direct  derivation of the conformal anomaly in the generalized 6D form factor in the mixed $x/p$-space representation \p{Fx}. It is an alternative to the Lagrangian insertion procedure from Sect.~\ref{s23}.

We start with the Fourier transform of the three-point function \p{2.2}, where we have restored the $i\ep$ prescriptions,
\begin{align}\label{b1}
I(x_1,x_2,p) := \int \frac{d^6 x_0}{i \pi^3} \frac{e^{i p x_0 }}{(x_{10}^2-i\ep)^2(x_{20}^2-i\ep)^2 (p^2+i\ep) } \,.
\end{align}
 When $p^2 \neq 0$ this integral is conformal,
\begin{align} \label{Ixp}
\left( \sum_{i=1}^2 K^{(x_i)}_{\mu;\Delta=2}  + K^{(p)}_{\mu;\Delta=2} \right) I(x_1,x_2,p) = 0 \,.
\end{align}
We want to understand what happens when $p^2 = 0$, that is, with the form factor \p{Fx}, 
\begin{align} \label{b3}
F(x_1,x_2,p)   =
 \lim_{p^2 \to 0} p^2 I(x_1,x_2,p)  = \left. \int \frac{d^6 x_0}{i \pi^3} \frac{e^{i p x_0 }}{(x_{10}^2-i\ep)^2(x_{20}^2-i\ep)^2 }\right|_{p^2 = 0}\,. 
\end{align}

\subsection{Differential equation for the form factor}\label{appB1}

Let us examine the properties of the function $F(x_1,x_2,p)$ following from its Lorentz, dilatation and translation invariance. The off- and on-shell realization of the  translations, eqs.~\p{a1} and \p{Pq}, respectively, together with Lorentz invariance imply that $F(x_1,x_2,p)= e^{\frac{i}{2}p(x_1+x_2)}\,  \varphi(x^2_{12},(p x_{12}))$. Further, the scaling behavior of the integral in \p{b3} is given by $1/x^2_{12}$, the rest should be a function of the dimensionless variable $(p x_{12})$ only. We thus arrive at the following general ansatz
\begin{align}\label{b4}
F(x_1,x_2,p)   =   e^{\frac{i}{2}p(x_1+x_2)}\,  \frac{\varphi(\a)}{x_{12}^2} \,, \qquad \a := \frac1{2} (p x_{12})\,, \quad p^2=0\,.
\end{align}
In addition, the function $\varphi(\a)$ must be even, 
\begin{align}\label{b4'}
\varphi(\a) = \varphi(-\a)\,,  
\end{align}
as a consequence of the permutation symmetry of the integral in \p{b3}. 

Now, hitting the integral by, e.g., $\Box_1$ and using the identity \cite{Gelfand}
\begin{align}\label{b.6}
\Box \frac1{(x^2-i\ep)^2}  = -4i\pi^3  \delta^{(6)}(x)\,, 
\end{align}
we find the inhomogeneous differential equation
\begin{align}\label{b6}
\Box_1   F(x_1,x_2,p)   = -4 \frac{e^{i p x_1 }}{x_{12}^4}\,.
\end{align}
Next, we use translation invariance to fix the frame $x_2=0$. Then the ansatz \p{b4} becomes
\begin{align}\label{B.8}
 F(x,0,p) = \frac1{x^2} e^{i\a}\,  {\varphi(\a)} := \frac1{x^2} f(\a) \,. 
\end{align}
  The differential equation  \p{b6} takes the form
\begin{align}\label{}
\Box \left[ \frac1{x^2} f(\a)  \right]  = -\frac4{x^4} [f(\a)+\a f'(\a) ] = - \frac{4}{x^4}\, e^{2i \a }\,.
\end{align}
Its solution  is $f(\a)= (e^{2i\a}-C)/(2i\a)$ with a constant $C$. Then $\varphi(\a)= (e^{i\a} - C e^{-i\a})/(2i\a)$ and the boundary condition \p{b4'} fixes $C=1$. We obtain the expression for the form factor
\begin{align}\label{b8}
F(x_1, x_2, p) 
  = \frac{1}{x_{12}^2}\, e^{\frac{i}{2}p(x_1+x_2)}\,  \frac{\sin\a}{\a}   \quad {\rm with} \  \ \a=\frac12 (p x_{12})\ ,
\end{align}
coinciding with \p{25}. Notice the absence of the unphysical pole at $(p x_{12})=0$. 

Later on we will also need the integral \p{b3} with modified prescription of the first propagator, 
\begin{align} \label{b8'}
\hat  F(x_1,x_2,p)   =
 \left. \int \frac{d^6 x_0}{i \pi^3} \frac{e^{i p x_0 }}{(x_{10}^2+i\ep)^2(x_{20}^2-i\ep)^2 }\right|_{p^2 = 0}\,. 
\end{align}
We rewrite the ansatz \p{b4} with a new function $\hat \varphi(\a)$. It satisfies a different boundary condition. The integral \p{b8'} changes sign under the simultaneous complex conjugation, exchange $x_1 \leftrightarrow x_2$ and also $p \leftrightarrow -p$. This operation leaves $\a$ invariant, therefore we require
\begin{align}\label{b9'}
\hat\varphi(\a) = -\hat\varphi(\a)^*\,. 
\end{align}
Next, we hit the integral  \p{b8'} with $\Box_1$ and use the complex conjugate of the identity \p{b.6}. Repeating the steps of fixing the frame $x_2=0$, defining $\hat  f(\a) = e^{i\a} \hat \varphi(\a)$, we obtain the differential equation
\begin{align}\label{}
\Box \left[ \frac1{x^2} \hat  f(\a)  \right]  = -\frac4{x^4} [\hat  f(\a)+\a \hat  f'(\a) ] = 4 \frac{e^{2i \a }}{x^4}\,.
\end{align}
Its solution is $\hat  f(\a)= (-e^{2i\a}-C)/(2i\a)$, hence $\hat \varphi(\a)= (-e^{i\a} - C e^{-i\a})/(2i\a)$. The boundary condition \p{b9'} fixes $C=1$ and we obtain $\hat \varphi(\a)= i\cos\a/\a$. Finally,
\begin{align}\label{b14}
\hat  F(x_1, x_2, p) 
  = \frac{i}{x_{12}^2}\, e^{\frac{i}{2}p(x_1+x_2)}\,  \frac{\cos\a}{\a} \,.
\end{align}
Note that this unphysical quantity has a pole at $(p x_{12}) = 0$. We define it with the principal value prescription, which is compatible with the boundary condition \p{b9'}.

\subsection{Conformal Ward identities}\label{appB2}

Let us examine the conformal properties of the result \p{b8}. We need to apply the off-shell $x-$space generator $K^{(x_i)}_{\mu;\Delta=2}$ (see \p{a2}) and the on-shell $p-$space generator $\mK^{(p)}_{\mu}$ (see \p{a12}). To this end it is sufficient to use translation invariance and fix the frame $x_2=0$.\footnote{The origin $x=0$ is stable under conformal transformations.} We go back to the general ansatz \p{B.8} and obtain
\begin{align}\label{b15}
 \left(  K^{(x)}_{\mu;\Delta=2} + K^{(p)}_{\mu;\Delta=4} \right)   \frac{f(\a)}{x^2}  = \frac{p_\mu}{4}\, \left[2i f' - f'' \right] +\frac{x_\mu}{x^2}\, \left[\a f'' + 2(1-i\a) f' -2i f  \right]\,.
\end{align}
Conformal invariance would mean that the coefficients of the two  vectors $p_\mu$ and $x_\mu$ vanish. These two equations are incompatible, so we conclude that  there is no function of the form \p{b4} satisfying the homogeneous Ward identity, i.e. which is an exact  conformal invariant. 

Now we take $f(\a)= (e^{2i\a}-1)/(2i\a)$ for the form factor \p{b8} and insert it in \p{b15}. The term $\sim x_\mu$ vanishes, while the term $\sim p_\mu$ yields the {anomalous Ward identity}  \p{26},
\begin{align}\label{b9}
\left( \sum_{i=1}^2 K^{(x_i)}_{\mu;\Delta=2} + K^{(p)}_{\mu;\Delta=4} \right) \, F  = p_\mu\, e^{\frac{i}{2} p (x_1 + x_2)} \frac{1}{2\a} \frac{d}{d\a} \frac{\sin\alpha}{\alpha} =: p_\mu A(x_1,x_2,p) \, ,
\end{align}
where we have restored the translation invariant dependence on $x_2$. 
Notice that if we know in advance that  the anomaly term is $\sim p_\mu$ only, as it is the case in \p{b9}, then the differential equation following from the vanishing of the term $\sim x_\mu$,  and the boundary condition \p{b4'} fix the solution \p{b8}, up to an overall constant.

The anomaly function $A(x_1,x_2,p)$  can also be rewritten in the integral form (see \p{27})
\begin{align} \label{b17'}
A(x_1,x_2,p) =  - \int^1_0 d\xi \,  \xi\bar\xi\, e^{ i (p x_{1})\xi +i (p x_2) \bar \xi}    \,.
\end{align}
This identity is easy to check, but it is also useful to interpret it as a Fourier transform. We again use translation invariance to fix $x_2=0$ and denote $\om \equiv (px_1)$. We get from \p{b9} 
\begin{align}\label{b18}
A(x_1,0,p) = \frac{e^{i\om}+1}{\om^2} +2i  \frac{e^{i\om}-1}{\om^3}\,.
\end{align}
Then we use the Fourier transform  \cite{Gelfand}\,\footnote{Here $\om^{-m}$ is defined as the finite part of the singular distribution.}
\begin{align}\label{b19}
\int_{-\infty}^\infty d\om\,  \om^{-m} \, e^{-i\om\xi} = \frac{(-i)^m\pi}{(m-1)!}\, {\rm sign}(\xi)\qquad {\rm for} \ m\geq 1 
\end{align}
and its inverse to rewrite \p{b18} in the integral form
\begin{align}\label{b20}
A(x_1,0,p) = - \int^1_0 d\xi \,  \xi\bar\xi\, e^{ i \om\xi}\,.
\end{align}
Restoring the point $x_2$ by translation invariance, we obtain   \p{b17'}.

In a similar way we derive the anomalous Ward identity for the integral $\hat{F}$ \p{b8'}, 
\begin{align}\label{b17}
\left( \sum_{i=1}^2 K^{(x_i)}_{\mu;\Delta=2} + K^{(p)}_{\mu;\Delta=4} \right) \, \hat{F}  = p_\mu\, e^{\frac{i}{2} p (x_1 + x_2)} \frac{i}{2\a} \frac{d}{d\a} \frac{\cos\alpha}{\alpha} =: p_\mu \hat{A}(x_1,x_2,p) \, .
\end{align}
The sum of the two anomalies \p{b9} and \p{b17} gives
\begin{align}\label{}
A + \hat{A} =  2e^{ipx_2}\, \frac{(px_{12}) -2i}{(px_{12})^3}\,.
\end{align}
Repeating the steps from \p{b18} to \p{b20},  we can rewrite this sum as follows:
\begin{align} \label{Ahat}
A + \hat{A} = -\int^{\infty}_{-\infty} d \xi\, \xi\bar\xi\, {\rm sign}(\xi)\,  e^{ i (p x_{1})\xi +i (p x_2) \bar \xi}\,.
\end{align}

\subsection{Evaluation by Schwinger parameters}\label{appB3}

Here we present another independent calculation of the form factor and its conformal anomaly. We first do the $d^6 x_0$ integration in \p{b1}, for $p^2\neq0$, by introducing Schwinger parameters. The result is a single-parameter integral  
of a modified Bessel function of the second kind \cite{Gelfand}.
Then we use the asymptotics for $p^2 \to 0$ and integrate over $\xi$,   
\begin{align}  \label{Iasymp}
I(x_1,x_2,p) &= \frac{e^{i p x_1}}{\sqrt{p^2 x_{12}^2}} \int^1_0 d\xi\, \sqrt{\xi\bar\xi}\  e^{-i \xi (p x_{12} )} 
K_1\left(\sqrt{x_{12}^2 p^2 \xi\bar\xi}\right)  =  \frac{\tau}{p^2} - \frac1{4} A \log p^2 + \ldots  
\end{align}
The integral has a pole $1/p^2$ and a cut starting at $p^2 = 0$.
The dots denote terms which are not singular  after acting with the conformal boost  \p{Kq} (e.g., the next term in the  expansion $p^2 \log p^2$ is finite after acting on it with $K^{(p)}_\mu$).
The residue  $\tau$ of the pole  is
\begin{align}\label{2.7}
 &\tau(x_1, x_2, p) =    \frac{1}{x_{12}^2}e^{\frac{i}{2}p(x_1+x_2)}\,  \frac{\sin\a}{\a} \quad {\rm with} \  \ \a=\frac12 (p x_{12})\,, \quad p^2\neq0\ .
 \end{align}
The expression for $A(x_1, x_2, p)$ is the same as in eq.~\p{b9} but   with $p^2 \neq 0$.

In the on-shell limit \p{b3} only the first term in \p{Iasymp} survives, and we find the generalized form factor  defined in \p{b3}, in accord with \p{b8},
\begin{align}\label{28}
F(x_1,x_2,p) = \tau(x_1, x_2, p^2 = 0) \,.
\end{align}

As we already know, this result is not invariant under conformal boosts. To evaluate the anomaly, we use the fact  that the off-shell integral $I(x_1,x_2,p)$ is conformal, so substituting the asymptotic expansion \p{Iasymp} in the Ward identity \p{Ixp} gives
\begin{align}
%%&K_\mu I(x_1,x_2,p) =0 \quad \Rightarrow   \nt
& \left( \sum_{i=1}^2 K^{(x_i)}_{\mu;\Delta=2} + K^{(p)}_{\mu;\Delta=2} \right) 
\frac{\tau}{p^2}  = \left( K^{(x_1)}_{\mu;\Delta=2} + K^{(x_2)}_{\mu;\Delta=2} + K^{(p)}_{\mu;\Delta=2} \right)  \frac1{4} A \log p^2 + \ldots  \,,
\end{align}
where the dots stand for the omitted nonsingular terms. Then we multiply both sides of this relation by $p^2$, 
take into account \p{a8} in the form
\begin{align} \label{interw}
&K^{(p)}_{\mu;\Delta = 2}\, 1/p^2\, \varphi(p) = 1/p^2 \, K^{(p)}_{\mu;\Delta=4} \, \varphi(p)\,,
\end{align}
and that $\log p^2$ produces a pole $1/p^2$ upon differentiation, 
\begin{align}
\lim_{p^2 \to 0} p^2 \, K^{(p)}_{\mu;\Delta=2}\, \log p^2 \, \varphi(p) = 4p_{\mu} \, \varphi(p)\,.
\end{align}
This enables us to  take the limit $p^2 \to 0$, 
\begin{align} \label{KI=A}
&\lim_{p^2 \to 0} \left( \sum_{i=1}^2 K^{(x_i)}_{\mu;\Delta=2} + K^{(p)}_{\mu;\Delta=4} \right) \, \tau(x_1,x_2,p)  =  p_\mu\, A(x_1, x_2, p^2 = 0) \, ,
\end{align}
in accord with \p{b9}. We see that the expression for the conformal anomaly is given by  the discontinuity $A(x_1,x_2,p)$ on the cut of the integral $I(x_1,x_2,p)$, eq.~\p{Iasymp}.

\section{Derivation of the Ward identity \p{55}}\label{s321}

\begin{figure}
\includegraphics[width = \textwidth]{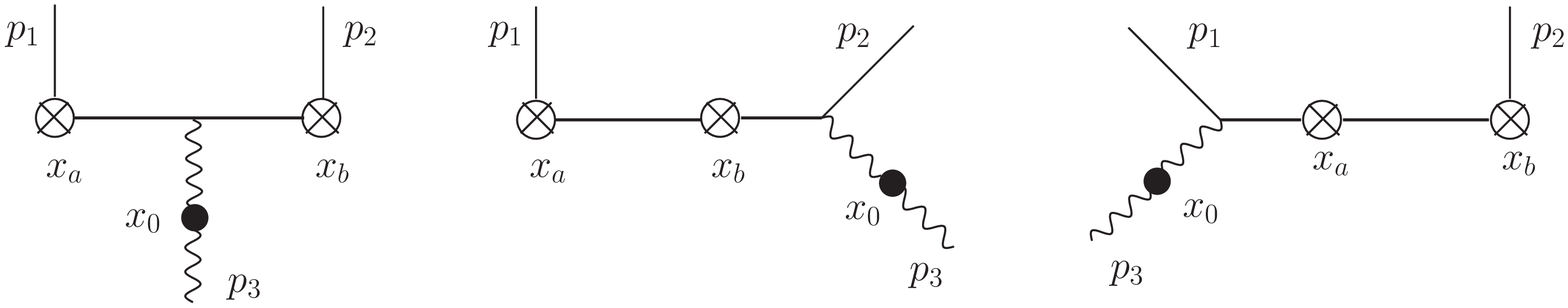}
\caption{Feynman graphs for the generalized form factor with Lagrangian insertion \p{FFYMLins}.} \label{FigYMFFLins}
\end{figure}
Here we show how the anomalous Ward identity \p{55} can be derived via the Lagrangian insertion method described in Section~\ref{s23}.
We need to find the residue  
\begin{align} \label{resYM}
\lim_{\ep \to0} \ep \int d^D x_0\, \vev{\cO(x_a) \cO(x_b) \left(  x_0^\mu \, L_{\rm YM}(x_0) \right) |\phi(p_1) \phi(p_2) g^{(+)}(p_3)}_{\rm Born}\,,
\end{align}
where $D= 4 - 2\ep$ and we insert the Yang-Mills part of the Lagrangian \p{Lagr}. 
We start with the generalized form factor of three operators with three on-shell particles. It is 
calculated in momentum space.
The contributing diagrams are shown in Fig.~\ref{FigYMFFLins}.
The result is   
\begin{align}\label{FFYMLins}
&\vev{\cO(q_a) \cO(q_b) L(q_0)|\phi(p_1) \phi(p_2) g^{(+)}(p_3)}_{\rm Born}\nt
&= \delta^{(4)}(P) \left[ \frac{[3|\tilde q_{1,a} q_{2,b}|3]}{q_{1,a}^2 q_{2,b}^2 q_{1,2,a,b}^2} 
+ \frac{\bra{2}q_{1,a,b}|3][23]}{q_{1,a}^2 q_{1,a,b}^2 q_{1,2,a,b}^2}
- \frac{\bra{1}q_{2,a,b}|3][13]}{q_{2,b}^2 q_{2,a,b}^2 q_{1,2,a,b}^2} \right] + {\rm perm}\ (a\leftrightarrow b)\,,
\end{align}
where $P$ is the total momentum, $P = q_a + q_b + p_1 + p_2 +p_3$.
Then we Fourier transform this result to coordinate space, $ q_{a},q_{b}, q_{0} \to x_{a}, x_{b}, x_{0}$. 
To transform the last two terms in \p{FFYMLins} we use \p{Fourier1}, and for
the first term we use the formula 
\begin{align} \label{Fourier2}
\int \frac{d^4 p}{4 \pi^2} \frac{d^4 q}{4 \pi^2} \frac{e^{- i p x- i q y} \vev{ \ell|p \,\tilde q|\ell}}{p^2 q^2 (q+p)^2} 
= -\frac{\vev{\ell| x\, \tilde y |\ell}}{x^2 y^2 (x-y)^2}\,.
\end{align}
The result for the generalized form factor in the mixed $x/p$-space representation is 
\begin{align}\label{YMFFx}
&\vev{\cO(x_a) \cO(x_b) L(x_0)|\phi(p_1) \phi(p_2) g^{(+)}(p_3)}_{\rm Born}=  
\frac{[3|\tilde x_{a0} x_{b0}|3]}{x^2_{a0} x^2_{b0} x^2_{ab}} \, e^{ix_a p_1 + i x_b p_2 + ix_0 p_3 } \nt 
&+
\frac{\bra{2}x_{b0}|3][23] }{\bra{2}x_{b0}|2]x^2_{ab} x^2_{b0}} \, e^{ix_0 p_3 + i x_a p_1 }\left(e^{ix_0 p_2} - e^{ix_b p_2}  \right) + 
\frac{\bra{1}x_{a0}|3][13] }{\bra{1}x_{a0}|1]x^2_{ab} x^2_{a0}} \, e^{ix_0 p_3 + i x_b p_2 }\left(e^{ix_a p_1} - e^{ix_0 p_1}  \right) \nt
 & +  {\rm perm}\ (a\leftrightarrow b)\,.
\end{align}
Like in \p{93}, the poles at $(x_{a0} p_1) = 0$ and $(x_{b0} p_2) = 0$  are absent.

Then we  substitute \p{YMFFx} in  \p{resYM} and extract the residue of the integral at the pole $1/\ep$.
We are interested in the divergent part, so we can use the $D=4$ integrand.
The last two terms in \p{YMFFx} do not contribute to the residue, while for the first term we obtain
\begin{align}
\int d^D x_0 \, x_0^\mu e^{i x_0 p_3} \frac{[3|\tilde x_{a0} x_{b0}|3]}{x^2_{a0} x^2_{b0} x^2_{ab}} =
\frac{i \pi^2}{2\ep} [3|\tilde\sigma^{\mu} x_{ab}|3] A (x_a,x_b,p_3) + O(\ep^0)\,,
\end{align}
with $A$ defined in \p{h4D}. Thus the Lagrangian insertion \p{resYM} yields the anomaly \p{55}.

\section{Unitarity cuts of the conformal Ward identities}\label{appD}

\begin{figure} 
\begin{align*}
\left( \sum_{i= 1}^3 \mK^\mu_i + K^\mu_{4} \right) \begin{array}{c}\includegraphics[width = 3cm]{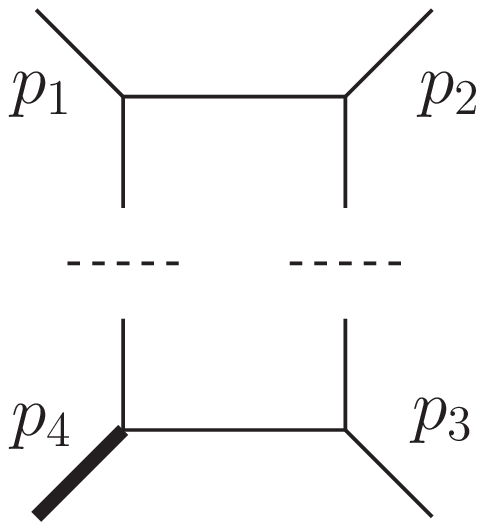}\end{array} & = \overbrace{\begin{array}{c}\includegraphics[width = 3cm]{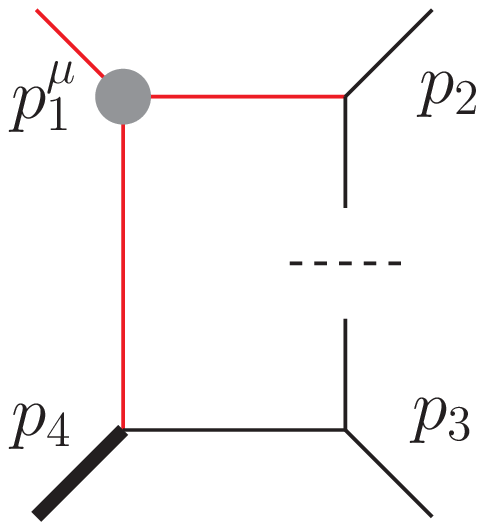}\hspace{-0.5cm}\end{array}}^{=0} + \overbrace{\begin{array}{c}\includegraphics[width = 3cm]{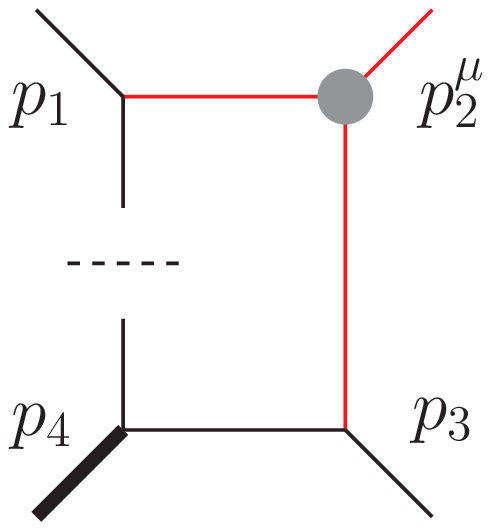}\hspace{-0.5cm}\end{array}}^{ = 0} + \begin{array}{c}\includegraphics[width = 3cm]{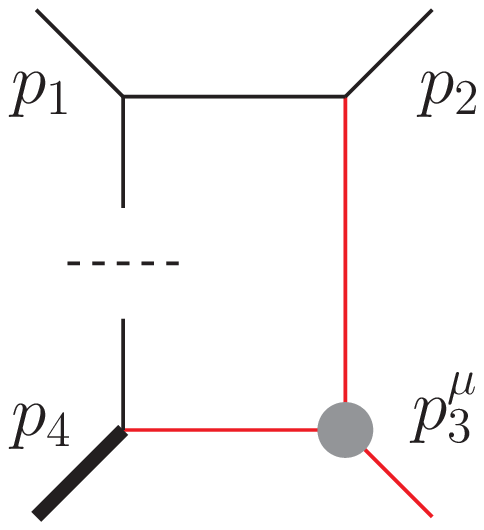}\end{array}
\end{align*}
\caption{Conformal Ward identity for the $s$-channel cut of the one-mass 6D box. The momenta of the highlighted lines are collinear. The first two contributions vanish.}
\label{fig1mCut}
\end{figure}
\begin{figure} 
\begin{align*}
\left( \mK^\mu_1 + K^\mu_{2} + \mK^\mu_3 + K^\mu_{4} \right) \begin{array}{c}\includegraphics[width = 3cm]{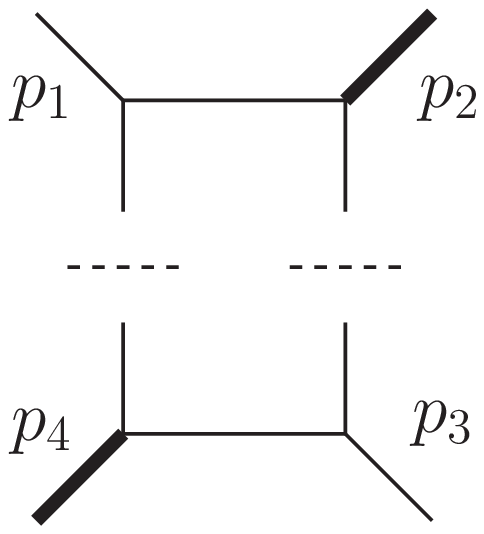}\end{array} & = \begin{array}{c}\includegraphics[width = 3cm]{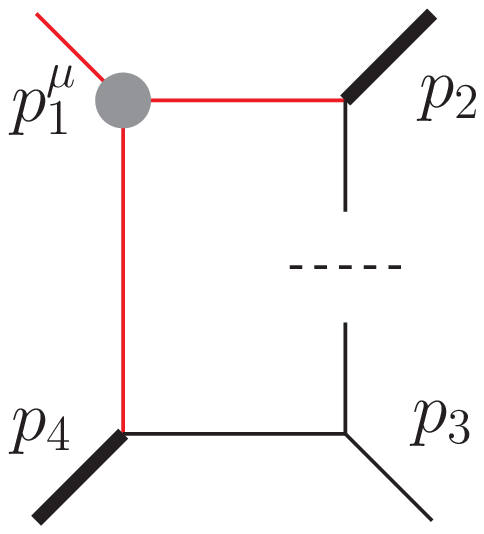}\end{array} + \begin{array}{c}\includegraphics[width = 3cm]{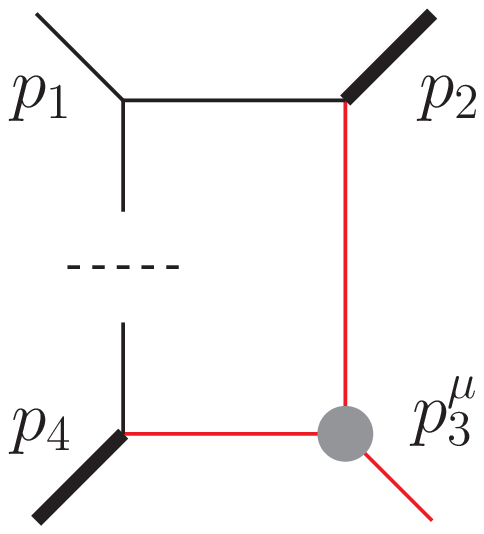}\end{array}
\end{align*}
\caption{Conformal Ward identity for the $s$-channel cut of the two-mass-easy 6D box.}
\label{fig2meCut}
\end{figure}

Here we discuss the cuts of the 6D boxes and the corresponding conformal Ward identities. 
For simplicity we consider the one-mass $I_{1m}$ and two-mass-easy $I_{2me}$ cases, Fig.~\ref{FigBoxInt}, 
and take the unitarity cut in the $s=p_{1,2}^2$ channel. The explicit expressions \p{3.3}, \p{34} have the cuts
\begin{align}\label{d1}
&\underset{p_{1,2}^2}{\rm Disc} \, I_{1m} = 2 \pi i \, \frac{1}{p_{1,3}^2} \left[ \log \left( \frac{p_{1,2}^2}{p_{2,3}^2} \right) + \log \left( 1- \frac{p_{4}^2}{p_{1,2}^2} \right)\right], \nt
&\underset{p_{1,2}^2}{\rm Disc} \, I_{2me} = 2 \pi i \, \frac{1}{p_{1,3}^2} \log \left( 1- a p_{1,2}^2 \right).
\end{align}
The integrals satisfy the conformal Ward identities \p{1mWI} and  \p{2meWI}, respectively, with the right-hand sides given explicitly by eqs.~\p{boxAnom'}. The anomalies are expressed in terms of logarithm functions, so they also have discontinuities in the variable $p_{1,2}^2$,
\begin{align}\label{d2}
&\underset{p_{1,2}^2}{\rm Disc} \left( A^{1\, \mu}_{\text{off/on}} + A^{2\, \mu}_{\text{on/on}} + A^{3\, \mu}_{\text{on/off}} \right) =
- 8 \pi i \, p_3^\mu  \, \frac{p_4^2}{p_{2,3}^2 (p_{3,4}^2 - p_4^2)^2} \,, \nt
& \underset{p_{1,2}^2}{\rm Disc} \left(  A^{1\, \mu}_{\text{off/off}} + A^{3\, \mu}_{\text{off/off}}  \right) =
\frac{8 \pi i\, p_{1,2}^2}{p_2^2 p_4^2 - p_{1,2}^2 p_{1,4}^2} \left[ \frac{p_1^\mu\, p_2^2}{(p_{1,2}^2 - p_2^2)^2} + \frac{p_3^\mu\, p_4^2}{(p_{1,2}^2 - p_4^2)^2} \right].
\end{align}

Now we can explicitly check that the cuts  \p{d1} of the integrals $I_{1m}$ and $I_{2me}$ satisfy conformal Ward identities whose right-hand sides are the cuts of the anomalies of the integrals \p{d2}: 
\begin{align}
&\left( \mK^\mu_1 + \mK^\mu_2 + \mK^\mu_3 + K^\mu_4\right) \delta^6(P)\, \underset{p_{1,2}^2}{\rm Disc} \, I_{1m} = \delta^6(P) \, \underset{p_{1,2}^2}{\rm Disc} \left( A^{1\, \mu}_{\text{off/on}} + A^{2\, \mu}_{\text{on/on}} + A^{3\, \mu}_{\text{on/off}} \right), \nt
&\left( \mK^\mu_1 + K^\mu_2 + \mK^\mu_3 + K^\mu_4\right) \delta^6(P)\,\underset{p_{1,2}^2}{\rm Disc}\, I_{2me} = \delta^6(P) \, \underset{p_{1,2}^2}{\rm Disc} \left( A^{1\, \mu}_{\text{off/off}} + A^{3\, \mu}_{\text{off/off}} \right) . \label{WIcut}
\end{align}
This is an example of  the generic phenomenon of commuting conformal anomaly and unitarity cut.

These conformal Ward identities for the discontinuities of the loop integrals follow from the anomaly of the cut of the trivalent vertex with $p^2 =0$ and with the loop momentum $\ell$ (see the proof below), 
\begin{align}\label{d4}
\left( K^{(\ell)}_{\mu;\Delta=2} + {\mK}^{(p)}_{\mu} \right) 
(-2 \pi i )\delta(\ell^2)\frac{1}{(\ell+p)^2}&   = 4 i \pi^3\, p_{\mu}\int^{\infty}_{-\infty} d \xi\, \xi\bar\xi\, {\rm sign}(\xi)\, \delta^{(6)}(\ell + \xi p) \,.
\end{align}
Using this result we can  derive the Ward identities \p{WIcut} for $I_{1m}$ and $I_{2me}$ by cutting the relevant propagators, see Figs.\,\ref{fig1mCut} and \ref{fig2meCut}. The first two contributions on the right-hand side of Fig.~\ref{fig1mCut} vanish. For instance, in the first term $\ell \sim p_1$ on the support of the anomaly. Momentum conservation at the vertex with $p_2$ and the cut condition imply that the lightlike vectors $\ell+p_1$, $p_2$ and $\ell+p_1-p_2$ at the vertex  are collinear.  However, we assume that $p_1$ and $p_2$ are generic lightlike momenta.

{\it Proof of \p{d4}:} Consider the sum of the integrals \p{b3} and \p{b8'},
\begin{align}\label{d5}
F+\hat F=\int \frac{d^6 x_0}{i \pi^3} \frac{e^{i p x_0 }}{(x_{20}^2-i\ep)^2 }\left[\frac1{(x_{10}^2-i\ep)^2} + \frac1{(x_{10}^2+i\ep)^2}\right]  \,.  
\end{align}
Its Fourier transform gives the cut of the propagator $1/(q^2_1+i\ep)$ 
in the trivalent vertex, 
\begin{align}\label{}
F+\hat{F} \ \stackrel{FT}{\rightarrow}\  \frac{\delta^{(6)}(q_1+q_2+p)}{q^2_2 +i\ep} \left[ \frac{i}{q_1^2+i\ep} -    \frac{i}{q_1^2-i\ep}\right] = \frac{\delta^{(6)}(q_1+q_2+p)}{q^2_2 +i\ep}  2 \pi  \delta(q_1^2)\,.
\end{align}
The anomaly of the sum $F+\hat{F}$ was found in  \p{Ahat}. Its Fourier transform  is straightforward and we arrive at \p{d4}.


\newpage
\begin{thebibliography}{99}

%\cite{Alday:2010zy}
\bibitem{Alday:2010zy}
  L.~Alday, B.~Eden, G.~P.~Korchemsky, J.~Maldacena and E.~Sokatchev,
  ``From correlation functions to Wilson loops,''
  JHEP {\bf 1109} (2011) 123
  %doi:10.1007/JHEP09(2011)123
  [arXiv:1007.3243 [hep-th]].
  %%CITATION = doi:10.1007/JHEP09(2011)123;%%
  
%\cite{Eden:2010zz}
\bibitem{Eden:2010zz}
  B.~Eden, G.~P.~Korchemsky and E.~Sokatchev,
  ``From correlation functions to scattering amplitudes,''
  JHEP {\bf 1112} (2011) 002
  %doi:10.1007/JHEP12(2011)002
  [arXiv:1007.3246 [hep-th]].
  %%CITATION = doi:10.1007/JHEP12(2011)002;%%  





%\cite{Drummond:2007au}
\bibitem{Drummond:2007au}
  J.~M.~Drummond, J.~Henn, G.~P.~Korchemsky and E.~Sokatchev,
  ``Conformal Ward identities for Wilson loops and a test of the duality with gluon amplitudes,''
  Nucl.\ Phys.\ B {\bf 826} (2010) 337
 % doi:10.1016/j.nuclphysb.2009.10.013
  [arXiv:0712.1223 [hep-th]].
  %%CITATION = doi:10.1016/j.nuclphysb.2009.10.013;%%

  
%\cite{Cachazo:2004by}
\bibitem{Cachazo:2004by}
  F.~Cachazo, P.~Svrcek and E.~Witten,
  ``Gauge theory amplitudes in twistor space and holomorphic anomaly,''
  JHEP {\bf 0410} (2004) 077
  %doi:10.1088/1126-6708/2004/10/077
  [hep-th/0409245].
  %%CITATION = doi:10.1088/1126-6708/2004/10/077;%%

%\cite{Bargheer:2009qu}
\bibitem{Bargheer:2009qu}
  T.~Bargheer, N.~Beisert, W.~Galleas, F.~Loebbert and T.~McLoughlin,
  ``Exacting N=4 Superconformal Symmetry,''
  JHEP {\bf 0911} (2009) 056
  %doi:10.1088/1126-6708/2009/11/056
  [arXiv:0905.3738 [hep-th]].
  %%CITATION = doi:10.1088/1126-6708/2009/11/056;%%


%\cite{Korchemsky:2009hm}
\bibitem{Korchemsky:2009hm}
  G.~P.~Korchemsky and E.~Sokatchev,
  ``Symmetries and analytic properties of scattering amplitudes in N=4 SYM,''
  Nucl.\ Phys.\ B {\bf 832} (2010) 1
  %doi:10.1016/j.nuclphysb.2010.01.022
  [arXiv:0906.1737 [hep-th]].
  %%CITATION = doi:10.1016/j.nuclphysb.2010.01.022;%%


%\cite{Bargheer:2011mm}
\bibitem{Bargheer:2011mm}
  T.\,Bargheer, N.\,Beisert and F.\,Loebbert,
  ``Exact Superconformal and Yangian Symmetry of Scattering Amplitudes,''
  J.\ Phys.\ A {\bf 44} (2011) 454012
  %doi:10.1088/1751-8113/44/45/454012
  [arXiv:1104.0700 [hep-th]].
  %%CITATION = doi:10.1088/1751-8113/44/45/454012;%%

%\cite{Engelund:2012re}
\bibitem{Engelund:2012re}
  O.~T.~Engelund and R.~Roiban,
  ``Correlation functions of local composite operators from generalized unitarity,''
  JHEP {\bf 1303} (2013) 172
  %doi:10.1007/JHEP03(2013)172
  [arXiv:1209.0227 [hep-th]].
  %%CITATION = doi:10.1007/JHEP03(2013)172;%%

%\cite{Witten:2003nn}
\bibitem{Witten:2003nn}
  E.~Witten,
  ``Perturbative gauge theory as a string theory in twistor space,''
  Commun.\ Math.\ Phys.\  {\bf 252} (2004) 189
  %doi:10.1007/s00220-004-1187-3
  [hep-th/0312171].
  %%CITATION = doi:10.1007/s00220-004-1187-3;%%


%\cite{CaronHuot:2011kk}
\bibitem{CaronHuot:2011kk}
  S.~Caron-Huot and S.~He,
  ``Jumpstarting the All-Loop S-Matrix of Planar N=4 Super Yang-Mills,''
  JHEP {\bf 1207} (2012) 174
  %doi:10.1007/JHEP07(2012)174
  [arXiv:1112.1060 [hep-th]].
  %%CITATION = doi:10.1007/JHEP07(2012)174;%%

%\cite{Bullimore:2011kg}
\bibitem{Bullimore:2011kg}
  M.~Bullimore and D.~Skinner,
  ``Descent Equations for Superamplitudes,''
  arXiv:1112.1056 [hep-th].
  %%CITATION = ARXIV:1112.1056;%%

%\cite{Chicherin:2016fac}
\bibitem{Chicherin:2016fac}
  D.~Chicherin and E.~Sokatchev,
  ``$ \mathcal{N} $ = 4 super-Yang-Mills in LHC superspace part I: classical and quantum theory,''
  JHEP {\bf 1702} (2017) 062
  %doi:10.1007/JHEP02(2017)062
  [arXiv:1601.06803 [hep-th]].
  %%CITATION = doi:10.1007/JHEP02(2017)062;%%
  
%\cite{Chicherin:2016fbj}
\bibitem{Chicherin:2016fbj}
  D.~Chicherin and E.~Sokatchev,
  ``$ \mathcal{N} $ = 4 super-Yang-Mills in LHC superspace part II: non-chiral correlation functions of the stress-tensor multiplet,''
  JHEP {\bf 1703} (2017) 048
  %doi:10.1007/JHEP03(2017)048
  [arXiv:1601.06804 [hep-th]].
  %%CITATION = doi:10.1007/JHEP03(2017)048;%%

%\cite{Broadhurst:1993ib}
\bibitem{Broadhurst:1993ib}
  D.~J.~Broadhurst,
  ``Summation of an infinite series of ladder diagrams,''
  Phys.\ Lett.\ B {\bf 307} (1993) 132.
  %doi:10.1016/0370-2693(93)90202-S
  %%CITATION = doi:10.1016/0370-2693(93)90202-S;%%


%\cite{Drummond:2006rz}
\bibitem{Drummond:2006rz}
 J.~M.~Drummond, J.~M.~Henn, V.~A.~Smirnov and E.~Sokatchev,
  ``Magic identities for conformal four-point integrals,''
  JHEP {\bf 0701} (2007) 064
  %doi:10.1088/1126-6708/2007/01/064
  [hep-th/0607160].
  %%CITATION = doi:10.1088/1126-6708/2007/01/064;%%
 


%\cite{Bern:1992em}
\bibitem{Bern:1992em}
  Z.~Bern, L.~J.~Dixon and D.~A.~Kosower,
  ``Dimensionally regulated one loop integrals,''
  Phys.\ Lett.\ B {\bf 302} (1993) 299;  Erratum: [Phys.\ Lett.\ B {\bf 318} (1993) 649]
  %doi:10.1016/0370-2693(93)90469-X, 10.1016/0370-2693(93)90400-C
  [hep-ph/9212308].
  %%CITATION = doi:10.1016/0370-2693(93)90469-X, 10.1016/0370-2693(93)90400-C;%%

%\cite{Anastasiou:1999cx}
\bibitem{Anastasiou:1999cx}
  C.~Anastasiou, E.~Glover and C.~Oleari,
  ``Application of  negative dimension approach to massless scalar box integrals,''
  Nucl. Phys. B {\bf 565} (2000) 445
  %doi:10.1016/S0550-3213(99)00636-7
  [hep-ph/9907523].
  %%CITATION = doi:10.1016/S0550-3213(99)00636-7;%%

%\cite{Dixon:2011ng}
\bibitem{Dixon:2011ng}
  L.~J.~Dixon, J.~M.~Drummond and J.~M.~Henn,
  ``The one-loop six-dimensional hexagon integral and its relation to MHV amplitudes in N=4 SYM,''
  JHEP {\bf 1106} (2011) 100
  %doi:10.1007/JHEP06(2011)100
  [arXiv:1104.2787 [hep-th]].
  %%CITATION = doi:10.1007/JHEP06(2011)100;%%
 
 %\cite{DelDuca:2011ne}
\bibitem{DelDuca:2011ne}
  V.~Del Duca, C.~Duhr and V.~A.~Smirnov,
  ``The massless hexagon integral in D = 6 dimensions,''
  Phys.\ Lett.\ B {\bf 703} (2011) 363
  %doi:10.1016/j.physletb.2011.07.079
  [arXiv:1104.2781 [hep-th]].
  %%CITATION = doi:10.1016/j.physletb.2011.07.079;%%  
  
%\cite{CaronHuot:2012ab}
\bibitem{CaronHuot:2012ab}
  S.~Caron-Huot and K.~J.~Larsen,
  ``Uniqueness of two-loop master contours,''
  JHEP {\bf 1210} (2012) 026
  %doi:10.1007/JHEP10(2012)026
  [arXiv:1205.0801 [hep-ph]].
  %%CITATION = doi:10.1007/JHEP10(2012)026;%%  


%\cite{Nandan:2013ip}
\bibitem{Nandan:2013ip}
  D.~Nandan, M.~F.~Paulos, M.~Spradlin and A.~Volovich,
  ``Star Integrals, Convolutions and Simplices,''
  JHEP {\bf 1305} (2013) 105
  %doi:10.1007/JHEP05(2013)105
  [arXiv:1301.2500 [hep-th]].
  %%CITATION = doi:10.1007/JHEP05(2013)105;%%
  

%\cite{Drummond:2010cz}
\bibitem{Drummond:2010cz}
  J.~M.~Drummond, J.~M.~Henn and J.~Trnka,
  ``New differential equations for on-shell loop integrals,''
  JHEP {\bf 1104} (2011) 083
  %doi:10.1007/JHEP04(2011)083
  [arXiv:1010.3679 [hep-th]].
  %%CITATION = doi:10.1007/JHEP04(2011)083;%%

%\cite{Drummond:2008vq}
\bibitem{Drummond:2008vq}
  J.~M.~Drummond, J.~Henn, G.~P.~Korchemsky and E.~Sokatchev,
  ``Dual superconformal symmetry of scattering amplitudes in N=4 super-Yang-Mills theory,''
  Nucl.\ Phys.\ B {\bf 828} (2010) 317
  %doi:10.1016/j.nuclphysb.2009.11.022
  [arXiv:0807.1095 [hep-th]].
  %%CITATION = doi:10.1016/j.nuclphysb.2009.11.022;%%

%\cite{Drummond:2009fd}
\bibitem{Drummond:2009fd}
  J.~M.~Drummond, J.~M.~Henn and J.~Plefka,
  ``Yangian symmetry of scattering amplitudes in N=4 super Yang-Mills theory,''
  JHEP {\bf 0905} (2009) 046
  %doi:10.1088/1126-6708/2009/05/046
  [arXiv:0902.2987 [hep-th]].
  %%CITATION = doi:10.1088/1126-6708/2009/05/046;%%


%\cite{Chicherin:2017frs}
\bibitem{Chicherin:2017frs}
  D.~Chicherin, V.~Kazakov, F.~Loebbert, D.~M\"{u}ller and D.~l.~Zhong,
  ``Yangian Symmetry for Fishnet Feynman Graphs,''
  arXiv:1708.00007 [hep-th].

%\cite{Cheung:2009dc}
\bibitem{Cheung:2009dc}
  C.~Cheung and D.~O'Connell,
  ``Amplitudes and Spinor-Helicity in Six Dimensions,''
  JHEP {\bf 0907} (2009) 075
 % doi:10.1088/1126-6708/2009/07/075
  [arXiv:0902.0981 [hep-th]].
  %%CITATION = doi:10.1088/1126-6708/2009/07/075;%%

%\cite{Gelfand}
\bibitem{Gelfand}
I.~M.~Gelfand and G.~E.~Shilov, 
  ``Generalized functions. Vol. 1. Properties and operations,'' Academic Press. San Diego (1964).
  
\end{thebibliography}
\end{document}